\documentclass{article}
\usepackage[utf8]{inputenc}
\usepackage{amsmath}

\usepackage{mathtools}
\usepackage{amsfonts}
\usepackage{amssymb}
\usepackage{algorithm}
\usepackage{algpseudocode}
\usepackage{enumerate}
\usepackage{todonotes}

\usepackage{natbib}
\usepackage{url}
\usepackage{color}
\usepackage{graphicx}
\usepackage{epstopdf}
\usepackage{enumitem}
\usepackage{multirow}
\usepackage{bm}
\usepackage{verbatim}
\usepackage{float}
\usepackage[toc,page]{appendix}
\usepackage[onehalfspacing]{setspace}
\usepackage{adjustbox}
\usepackage{rotating}
\usepackage{subcaption}
\usepackage{amsthm}
\usepackage{booktabs,dcolumn,caption}
\usepackage{tikz}
\usepackage{authblk}
\usepackage{etoolbox}
\usepackage{lmodern}
\usepackage{hyperref}
\hypersetup{
    colorlinks=true,
    linkcolor=blue,
    citecolor=blue,
    filecolor=blue,      
    urlcolor=magenta,
    pdftitle={Overleaf Example},
    pdfpagemode=FullScreen,
    }

\graphicspath{{Fig/}}

\makeatletter
\patchcmd{\@maketitle}{\LARGE \@title}{\fontsize{14}{19.2}\selectfont\@title}{}{}
\makeatother

\theoremstyle{remark}

\numberwithin{equation}{section}

\title{Real-time small area estimation of food security in Zimbabwe:\\integrating mobile-phone and face-to-face surveys using joint multilevel regression and poststratification}

\author
{Sahoko Ishida$^{1}$, Adam Howes$^{1,2}$, Valerie Bradley$^{1}$, Elizaveta Semenova$^{3}$,  Theo Rashid$^{4}$,
Silinganisiwe Dzumbunu$^{5}$,
Sumali Bajaj$^{6}$,
Gaurav Singhal$^{7}$, Dino Sejdinovic$^{8}$,
Herbert Zvirere$^{9}$,
Duccio Piovani$^{10}$, Silvia Passeri$^{10}$,
Kusum Hachhethu$^{10}$,
Arif Husain$^{10}$,
George D. Kembo$^{11}$,
Terrence Kairiza$^{12}$,
Seth Flaxman$^{1\ast}$ 
\\
\small{$^{1}$Department of Computer Science, University of Oxford},\\
\small{$^{2}$Department of Mathematics, Imperial College London},\\
\small{$^{3}$Department of Epidemiology and Biostatistics, Imperial College London}\\
\small{$^{4}$Amazon},\\
\small{$^{5}$Accelerate Research Hub, Centre for Social Science Research, University of Cape Town}\\
\small{$^{6}$Merton College,
Department of Statistics \&
Pandemic Sciences Institute, University of Oxford},\\
\small{$^{7}$Arboreum.dev},\\
\small{$^{8}$School of Computer and Mathematical Sciences \& Australian Institute for Machine Learning, \\University of Adelaide},\\
\small{$^{9}$ World Food Programme Zimbabwe},\\
\small{$^{10}$ Analysis Performance and Planning Division, World Food Programme} \\
\small{$^{11}$Food and Nutrition Council Zimbabwe}\\
\small{$^{12}$Department of Economics, Bindura University of Science Education}
\\
\small{$^\ast$To whom correspondence should be addressed; E-mail:  seth.flaxman@cs.ox.ac.uk.}
}
\date{\vspace{-5ex}}
\begin{document}

\maketitle

\begin{abstract}
\noindent 
Real-time, fine-grained monitoring of food security is essential for enabling timely and targeted interventions, thereby supporting the global goal of achieving zero hunger—a key objective of the 2030 Agenda for Sustainable Development. Mobile phone surveys provide a scalable and temporally rich data source that can be tailored to different administrative levels.
However, due to cost and operational constraints, maintaining high-frequency data collection while ensuring representativeness at lower administrative levels is often infeasible.
We propose a joint multilevel regression and poststratification (jMRP) approach that combines high-frequency and up-to-date mobile phone survey data, designed for higher administrative levels, with an annual face-to-face survey representative at lower levels to produce reliable food security estimates at spatially and temporally finer scales than those originally targeted by the surveys.
This methodology accounts for systematic differences in survey responses due to modality and socio-economic characteristics, reducing both sampling and modality bias. We implement the approach in a fully Bayesian manner to quantify uncertainty.
We demonstrate the effectiveness of our method using data from Zimbabwe, thus offering a cost-effective solution for real-time monitoring and strengthening decision-making in resource-constrained settings.
\end{abstract}

\section{Introduction}\label{section:introduction}
Food security remains a critical global challenge. According to the latest State of Food Security and Nutrition in the World report \citep{fao2024}, approximately 733 million people experienced hunger in 2023, equivalent to one in eleven individuals globally and one in five in Africa. This emergency is driven by a complex interplay of factors, including climate and environmental shocks, economic instability, and political conflicts. As such, real-time, spatially granular monitoring of food security is essential for governments and organisations like the World Food Programme (WFP) to effectively plan aid, assistance, and preemptive actions to mitigate the impacts of food crises.

Traditionally, face-to-face (F2F) surveys have been the gold standard for estimating food security indicators. However, they are resource-intensive and often too costly and impractical for frequent data collection, especially in remote or unstable areas. 
In contrast, mobile phone surveys offer a more scalable and cost-effective alternative.
An example is the WFP's mobile-phone surveys, extensively used in the Mobile Vulnerability Analysis and Mapping (mVAM) project to monitor food security in areas with restricted humanitarian access. Following successful pilots in the Democratic Republic of Congo and Somalia, the project expanded, becoming operational in multiple countries and a vital tool for WFP, exemplified during the 2014 Ebola Virus Disease emergency response. The project subsequently evolved to incorporate a continuous, real-time monitoring system built on systematic daily data collection \citep{worldfoodprogramme2021real}. This transformation enhanced WFP's capacity to trigger timely interventions, mitigate risks, and respond more effectively to future shocks.

The design of mVAM surveys varies across countries based on operational priorities and resource constraints. These factors determine key aspects of survey implementation, including geographic coverage, frequency, and the administrative level at which the survey is designed to produce reliable estimates. In many cases, collecting sufficiently large samples at high frequency to cover all lower administrative areas is financially infeasible. As a result, some countries collect data at finer spatial scales (e.g., the second or third administrative level) only for specific areas rather than nationwide or conduct surveys at these levels less frequently. In other cases, surveys are designed for frequent data collection at the first administrative level to support high-level surveillance, while other data sources are used for local-level targeting. Given these constraints, developing methodologies that provide reliable lower administrative-level estimates from surveys originally designed for higher administrative-level analysis is beneficial, allowing for more efficient use of limited resources and enabling real-time food security monitoring at high spatial resolution.

Additionally, mobile-phone-based surveys are subject to two types of biases: sampling bias and modality bias. Mobile phone surveys reach only individuals with phone access, which can contribute to sampling bias if phone ownership is linked to socio-demographic characteristics that influence food security, potentially leading to over- or under-representation of specific population groups. To mitigate this, WFP applies poststratification using demographic and geographic weights derived from nationally representative surveys. Households are weighted based on key socio-demographic characteristics, such as education level, to better align the sample with the target population. However, when attempting to obtain estimates at finer spatial scales than the survey was designed for, poststratification becomes less straightforward due to limited sample sizes, reducing the ability to adjust for multiple variables. Modality bias refers to potential differences in how people respond to phone surveys compared to F2F interviews, which can affect the accuracy of the data collected. Addressing both biases is crucial to ensure the reliability of food security estimates.

\subsection{Overview of existing methodologies}
The small area estimation (SAE) literature provides methods for producing reliable estimates when sample sizes are insufficient for direct, design-based estimates at the desired level (see, e.g., \cite{ghosh1994small, pfeffermann2002small, pfeffermann2013new, rao2015small} for an overview). SAE methodologies have been further extended to incorporate spatial \citep{ghosh1994small, pratesi2008small} and spatio-temporal correlations \citep{mercer2015space, gao2024smoothed}.

When addressing sampling bias, the standard approach is poststratification \citep{little1993post}. It adjusts survey estimates by applying weights to respondents, ensuring that the weighted sample better reflects the target population based on auxiliary variables, such as demographic characteristics. An approximation to poststratification is raking \citep{deming1940least, deville1992calibration, skinner2017introduction}, an iterative algorithm that uses marginal totals to adjust for bias and can be used even when the joint distribution of auxiliary variables is unknown.  A more modern approach for correcting non-representative samples is multilevel regression and poststratification (MRP), first introduced by \cite{gelman1997poststratification}, with growing literature on both methodology \citep{gao2021improving, gelman2006data, ghitza2013deep, kuh2024using} and application \citep{downes2018multilevel, lax2009gay,wang2015forecasting,zhang2014multilevel}. MRP integrates hierarchical data modelling to borrow information across subgroups, adjusting for sample non-representativeness through poststratification. Since the estimates are computed as a weighted average of predictions for each subgroup, MRP does not suffer from the problem of empty cells, which is often encountered in simple poststratification methods when adjusting for many auxiliary variables. MRP also serves as a natural framework for the SAE problem \citep{zhang2014multilevel}. Recently \cite{gao2021improving} proposed using structured priors, such as autoregressive, or conditional auto-regressive (CAR) models \citep{besag1974spatial}, which allow for non-uniform information borrowing and help reduce bias in temporally or spatially referenced data. 

One approach to addressing modality bias is to incorporate multiple modes of data collection in the modelling process \citep{elliott2005obtaining, elliott2007use}. A recent study by \cite{gellar2023calibrated} introduced a calibrated MRP approach, which integrates joint modelling of multiple modalities within MRP to mitigate both modality bias and residual sampling bias. Residual sampling bias refers to the bias that remains unaccounted for in the standard MRP approach, which typically uses a single survey modality (e.g., a mobile-phone-based survey). This bias occurs when there are differences between the sample (in this case, from a mobile-phone survey) and the target population.

\subsection{Proposed methodology}
We propose a joint spatio-temporal MRP approach for real-time small area estimation of a food security indicator. By integrating survey data from multiple sources and modes, this methodology addresses modality bias, sampling bias, and the small sample size at the desired spatial and temporal resolution. 

Specifically, we leverage: (1) a high-frequency mobile phone survey conducted at the first administrative level, which provides real-time insights but lacks sufficient statistical power at finer spatial resolutions and is susceptible to sampling and modality bias, and (2) a low-frequency F2F survey, which is more representative of the target population at a granular spatial level but cannot provide frequent updates. Our approach uses multilevel regression models to control for survey modality and account for spatial and temporal autocorrelation through structured priors.

Poststratification is used to aggregate cell-level predictions to the desired spatial and temporal domains. When necessary, we apply raking to derive appropriate weights, ensuring representativeness with respect to the target population. For real-time monitoring, we compute joint MRP estimates across space and time. This can be viewed as a nowcasting effort, calibrating real-time mobile phone survey data to produce timely and population-representative estimates.

Our contribution is novel in applying a joint MRP framework to small area estimation in a spatio-temporal setting. While related methodologies have been proposed—for example, a calibrated MRP approach by \cite{gellar2023calibrated} and an MRP method with structured priors by \cite{gao2021improving}—to the best of our knowledge, our approach is the first to combine these elements within a single MRP framework in a spatio-temporal setting.
By integrating high-frequency mobile phone surveys, designed for higher administrative levels, with lower-frequency F2F surveys that are spatially representative at finer scales, our methodology generates reliable, frequently updated estimates at lower administrative levels while mitigating sampling and modality biases in mobile phone surveys. This offers a cost-effective solution for organisations and policymakers aiming to strengthen food security monitoring within resource constraints.

\subsection{Outline}
We demonstrate the effectiveness of our joint MRP methodology through a case study in Zimbabwe.
The remainder of this paper is structured as follows: Section \ref{section: motivation and data} presents a motivating example from Zimbabwe and introduces the dataset, including WFP’s mobile-phone-based survey and the annual F2F survey. Section \ref{section: method} details the methodology for estimating a food security indicator at a monthly cadence and the second administrative level. Section \ref{section: result} validates the approach and presents the results. Finally, Section \ref{section: conclusion} summarises the findings and discusses limitations and future research directions.



\section{Motivation: real-time monitoring of food security in Zimbabwe}\label{section: motivation and data}
Zimbabwe is classified as a “food-deficit” country and ranks among the lowest on the 2022 Global Hunger Index \citep{globalhungerindex}. Persistent climate and economic challenges, including recurrent droughts, floods, and high inflation, have contributed to deteriorating food security, necessitating continuous monitoring and assessment.

WFP monitors food security in Zimbabwe through the mVAM project and its real-time monitoring (RTM) system. The RTM system collects data daily via mobile phone surveys, providing timely insights into food security conditions. Since its inception, mVAM has been a critical tool for WFP, governments, and international organisations, particularly during crises such as the COVID-19 pandemic when F2F data collection was restricted.
The design of mVAM surveys varies by country, with spatial and temporal resolution determined by specific operational needs and financial constraints. In Zimbabwe, the survey is conducted to support real-time monitoring at the first administrative level. While finer spatial granularity (e.g., second administrative level) is desirable for identifying the most vulnerable populations and enabling more targeted interventions, achieving this at high frequency requires significantly greater resources. Further details on the RTM system and mVAM survey design are provided in Appendix \ref{appendix: mVAM data processing} and \cite{worldfoodprogramme2021real}.
\begin{figure}[t]
    \centering
    \includegraphics[width=0.99\linewidth]{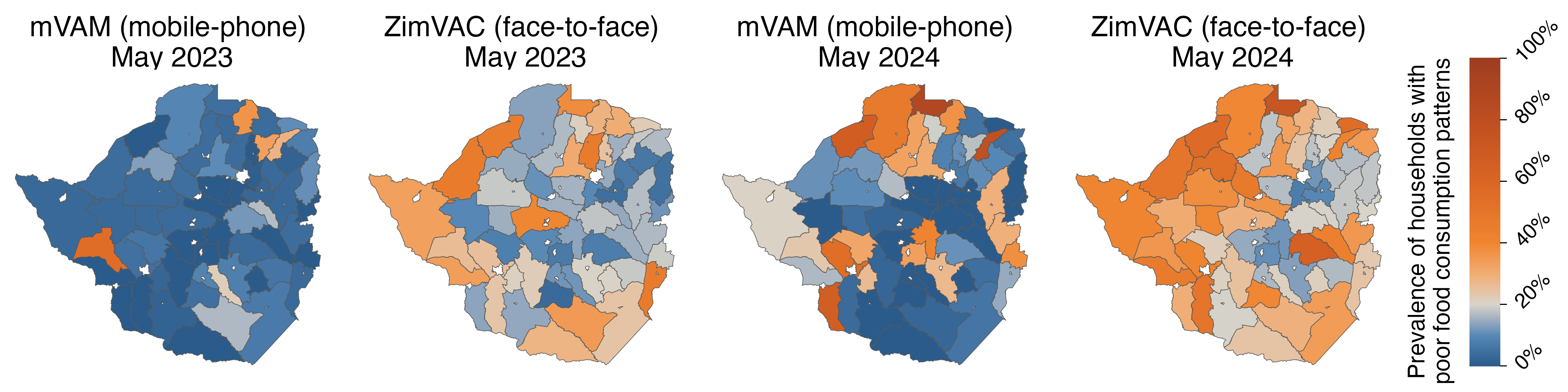}
    \caption{Comparison of direct estimates of the prevalence of households with poor food consumption patterns for May 2023 and May 2024. These estimates represent district-level prevalence in rural Zimbabwe. The mobile-phone survey (mVAM) is designed for province-level analysis in Zimbabwe, and its direct estimates are shown in the first and third panels, while the face-to-face survey (ZimVAC) direct estimates, representative at the district level, are displayed in the second and fourth panels. The colour scale (blue: $0$\% to orange: $100$\%) shifts at $20$\% to highlight prevalence considered at least moderately high.}
    \label{fig: mVAM-ZimVAC example}
\end{figure}


In addition to mVAM, Zimbabwe’s food security is monitored through an annual F2F survey conducted by the Zimbabwe Vulnerability Assessment Committee (ZimVAC). This is designed to be representative at the second administrative level, providing more granular insights on food security conditions. However, its lower frequency makes it less responsive to short-term fluctuations compared to the mVAM survey.

Figure \ref{fig: mVAM-ZimVAC example} illustrates the differences between mobile phone and F2F survey estimates at the second administrative level in Zimbabwe. In Zimbabwe, the first administrative level corresponds to provinces, while the second administrative level consists of districts. Since the mobile-phone survey is designed for analysis at a higher administrative level, it may fail to capture localised variations in food security. Moreover, differences in survey modality and sample composition can further introduce biases, potentially underestimating food security prevalence in certain areas. In contrast, the face-to-face survey, designed to be representative at the district level, offers a more granular perspective and reveals spatial variations in food security that mobile-phone-based direct estimates may overlook.

Food security is a complex concept measured through various indicators. One key indicator derived from the mVAM and ZimVAC survey data is the Food Consumption Score (FCS), which assesses household food consumption by measuring the frequency of consumption of eight different food groups over the past seven days. The FCS ranges from 0 to 112, with specific thresholds used to classify households into different levels of food security. In Zimbabwe, poor food consumption patterns are defined as an FCS of 28 or below. Further details on the FCS are provided in Appendix \ref{appendix: FCS}. While there are other indicators for food security, such as the reduced Coping Strategy Index (rCSI), which captures households' behavioural responses to food shortages, this paper focuses on FCS and aims to estimate the monthly prevalence of households with poor food consumption in Zimbabwe’s rural districts. We integrate real-time mobile phone surveys and annual F2F surveys, each with different temporal frequencies (daily and annually) and spatial granularities (province and district).
We use the following data sources:
\begin{itemize}
\item The mVAM survey in Zimbabwe: A daily mobile-phone-based survey covering September 2020 to June 2024, with 43,090 observations. Data collection was temporarily interrupted between March and April 2024. The survey is stratified by province and employs a complex rotation panel design to account for attrition and non-response of those with a mobile phone. In Zimbabwe, it aims to survey 150 households per province within a 30-day rolling analysis window.
\item The Rural ZimVAC survey: An annual (May) F2F survey designed to be representative at the district level, focusing on rural areas. The survey covered around 15,000 respondents in 2023 and 18,000 in 2024 across 60 rural districts \citep{zimbabwe_vulnerability_assessment_2023, zimlac2024}, ensuring a minimum sample size of 250 households per district in 2023 and 300 in 2024. Further details on the sampling scheme are provided in Appendix \ref{appendix: ZimVAC}. Although ZimVAC conducts a separate annual survey for urban populations, our analysis focuses on districts covered in the Rural ZimVAC survey, given their higher vulnerability to food security. Additionally, aggregated district-level data from the 2021 and 2022 ZimVAC surveys, collected using a different set of food items, are available and used for reference.
\end{itemize}
Beyond the data needed to calculate food security indicators such as FCS, the mVAM and ZimVAC surveys also collect demographic and socioeconomic information. Table \ref{table:covariates} lists a subset of these variables, specifically those used in our modelling. Additionally, we incorporate the following data sources for the poststratification step discussed in Section \ref{section: poststratification}:
\begin{itemize}
\item The 2015 Zimbabwe Demographic and Health Survey (2015 DHS): A survey covering 11,000 households across all provinces, providing demographic and health data representative at the national and provincial levels \citep{ZDHS2015}.
\item The 2022 Zimbabwe Population and Housing Census (2022 Census): Conducted by the Zimbabwe National Statistics Agency in April 2022, providing marginal counts for key variables of interest \citep{zimstat2022census}.
\end{itemize}

\section{Methods}\label{section: method}
In this section, we present our methodology for estimating the monthly district-level prevalence of households with poor food consumption patterns (FCS $\leq$ 28) by integrating data from two sources: the high-frequency mobile-phone-based mVAM survey, designed for province-level analysis in the case of Zimbabwe, and the annual F2F ZimVAC survey, which is representative at the district level. To facilitate joint modelling, we index mVAM observations by month rather than exact survey dates, aligning the temporal resolution of the two surveys.

We build on multilevel regression and poststratification (MRP) \citep{gelman1997poststratification}, adapted for small area estimation (SAE) \citep{rao2015small, pfeffermann2013new}. The first stage fits a joint Bayesian multilevel regression model to both survey datasets to estimate the probability of exhibiting poor food consumption patterns based on location, time, and household characteristics. Structured spatial and temporal priors enable partial pooling across districts and months, reducing variance. The use of such priors is well-established in both the SAE \citep{ghosh1994small, pratesi2008small, mercer2015space, gao2024smoothed} and MRP literature \citep{gao2021improving}. To address modality bias, we include a survey mode indicator, explicitly accounting for systematic differences between mobile phone and F2F surveys. While modelling multiple surveys with different modalities has been explored \citep{elliott2005obtaining, elliott2007use}, including recent work by \cite{gellar2023calibrated} within the MRP framework, its application in a spatio-temporal SAE context remains novel.

In the second stage, poststratification adjusts for sampling bias by weighting the modelled predictions according to population proportions across geographic, demographic, and socio-economic variables. 
Simple poststratification \citep{little1993post} often struggles with sparse or empty cells when many adjustment variables are included. MRP mitigates this issue through hierarchical modelling, allowing for information sharing across groups. We refer to the resulting modelled estimates as joint MRP (jMRP). The final outputs are monthly district-level estimates of poor food consumption prevalence, integrating both data sources while quantifying uncertainty.

\subsection{Joint multilevel regression model with spatio-temporal smoothing}\label{section: model}
Our model, which is a generalised linear mixed model (i.e.~multilevel logistic regression), has fixed effects of household-level covariates and spatial and temporal random effects. The outcome variable $y$ is a indicator binary variable: $y=1$ if the FCS $\leq$ 28 and $0$ if FCS $> 28$. We model the probability $p_i$ that a given household $i$ is in the poor food consumption group by
\begin{equation}\label{eq: main logit model}
   \log{\left(\frac{p_i}{1-p_i}\right)} = \gamma + \mathbf{x}_i^\top \boldsymbol{\beta} + \phi_{s[i]} + \zeta_{s[i]} + \nu_{t[i]}+ \xi_{t[i]} + \psi_{r[i]t[i]} , 
\end{equation}
where $\gamma$ is the intercept term, $\mathbf{x}_i$ is the vector of covariates, and $ \phi_{s[i]}, \zeta_{s[i]},\nu_{t[i]}$, $\xi_{t[i]}$ and $\psi_{r[i]t[i]}$ are random effect terms which we describe below. The indices $s,r,t$ are for district, province and month of the survey as detailed in Table \ref{table:covariates}. In addition to our main joint model, we also consider a mobile-phone-only model $\mathcal{M}_\text{MPonly}$, for comparison. We fit the generalised linear model with six covariates: \textit{Water source, Head of household education, Female head of household, Household size, Household toilet type} only to the mVAM survey. These covariates are also listed $2-7$ in Table \ref{table:covariates}. 
\begin{table}[t]
\caption{Summary of indices and covariates used in the regression models: $\mathcal{M}_\text{MPonly}$ and $\mathcal{M}_\text{Joint}$ referring to the mobile-phone-only model and joint model, respectively. The table lists the indices, their descriptions, and the relevant covariates, including demographic, socio-economic, and survey modality variables. Interaction terms between the survey modality and other covariates are included in the joint model. Checkmarks indicate whether the respective covariate is included in each regression model.}
\label{table:covariates}
\begin{center}
\begin{tabular}{ p{5.25cm}p{6.0cm}p{1.125cm}p{1.125cm}}
  \hline
  \textbf{Index} & \multicolumn{3}{l}{\textbf{Description}} \\
  \hline
  Household (indexed by $i$) & \multicolumn{3}{p{8.25cm}}{$i = 1, 2, \ldots, N$.  For our data analysis, $N=43,090$ for the mobile-phone-only model and $N=57,655$ for the joint model.}  \\
  District (indexed by $s$) & \multicolumn{3}{p{8.25cm}}{$s = 1, 2, \ldots, S$ For our data analysis $S=60$, reflecting the number of districts in Rural ZimVAC study.} \\
  Province (indexed by $r$) & \multicolumn{3}{p{8.25cm}}{$r = 1, 2, \ldots, R$. For our data analysis $R=8$, reflecting the number of provinces in Zimbabwe.}\\
  Month of survey (indexed by $t$) & \multicolumn{3}{p{8.25cm}}{$t = 1, 2, \ldots, T$. For our data analysis, $T=46$ and the study period ranges from September 2020 to June 2024.}
  \\ \hline 
 \textbf{Covariate} & \textbf{Values} & \multicolumn{1}{c}{$\mathcal{M}_\text{MPonly}$} &\multicolumn{1}{c}{$\mathcal{M}_\text{Joint}$} \\\hline
  1: Survey modality & face-to-face (F2F) / mobile phone  & & \multicolumn{1}{c}{\checkmark}\\
  2: Water source & improved/other & \multicolumn{1}{c}{\checkmark} & \multicolumn{1}{c}{\checkmark}\\
  3: Head of household education & none, primary, secondary, higher & \multicolumn{1}{c}{\checkmark} & \multicolumn{1}{c}{\checkmark}\\
  4: Male head of household & yes/no & \multicolumn{1}{c}{\checkmark} & \multicolumn{1}{c}{\checkmark} \\
  5: Household size & 1-2 people, 3-4, 5-6, 7+ people & \multicolumn{1}{c}{\checkmark} & \multicolumn{1}{c}{\checkmark}\\
  6: Household toilet type & improved/unimproved & \multicolumn{1}{c}{\checkmark} & \multicolumn{1}{c}{\checkmark}\\
  7: Household mobile phone ownership & yes/no (ZimVAC), probability (mVAM)& \multicolumn{1}{c}{\checkmark} & \multicolumn{1}{c}{\checkmark} \\
  \multicolumn{2}{l}{8-12: Interaction between 1 (Modality) and each of  2-6}& & \multicolumn{1}{c}{\checkmark}\\ \hline
\end{tabular}
\end{center}
\end{table}
For the joint model $\mathcal{M}_\text{Joint}$, we additionally control for the modality of the survey (mobile phone/F2F) and incorporate interaction effects between this variable and each of the 6 covariates ($2-7$ in the table).

One of the key variables included in both models is household mobile phone ownership. Naturally, for the mobile-phone-only model, this variable has the same value (yes, or $1$) for all households, meaning it should be absorbed into the intercept term. However, mobile-phone sampling introduces a source of sampling bias, making it an important and informative covariate. To address this, for households in the mobile-phone survey, we impute the household mobile phone ownership variable by estimating each household's probability of owning a mobile phone, given a set of socio-economic and demographic variables. Specifically, to estimate the probability of phone ownership for each household in the mobile-phone survey, we fit a multilevel logistic regression to either the 2019 Multiple Indicator Cluster Survey (MICS) data (for the mobile-phone-only model) or the ZimVAC 2023 data (for the joint model) and compute the fitted values for all households in the mobile-phone survey.

Our model \eqref{eq: main logit model} includes two district-level random effects corresponding to the convolution model of \cite{besag1991bayesian}. The term $\zeta_s$ models the unstructured component, which is assumed to follow an independent and identically distributed (i.i.d.) normal distribution, i.e., $\zeta_s\sim \mathcal{N}(0,\sigma_\zeta^2)$ for $s=1,\ldots, S$.
The intrinsic conditional autoregressive (ICAR) terms $\phi_s$ model a structural spatial component. Conditional on all the other $\phi_l, l\neq s$, the term $\phi_s$ follows a normal distribution $N(\sum_{j\in\partial_s}\frac{\phi_j}{d_s},\frac{\sigma_\phi}{d_s})$ where $\partial_s$ denotes the set of neighbours for the regions $s$, and $d_s = |\partial_s|$ is the number of neighbours of region $s$. See \cite{rue2005gaussian} for the details of the ICAR model. It is common practice to constrain $\sum_{s=1}^{S} \phi_s = 0$, but we replace this with a soft sum-to-zero constraint to keep the sum as close to zero as possible; see \cite{morris2019bayesian}.

We take a similar approach for temporal random effect terms $\nu_t$ and $\xi_t$. To model the temporal dependence, we assume the temporal random effect term $\nu_t$ to follow a first-order random walk, $\nu_t|\nu_{t-1}\sim \mathcal{N}(\nu_{t-1},\sigma^2_\nu)$. This is combined with an unstructured random effect term $\xi_t$ which follows an i.i.d normal distribution $N(0,\sigma^2_\xi)$ and can capture short-term variations and deviations from the random walk process. 

We allow variation in the time trend at the province level by including the spatio-temporal interaction term: $\psi_{rt}\sim \mathcal{N}(0,\sigma_\psi)$ for $r=1,\ldots R$ and $t=1,\ldots, T$. This reflects the assumption that unstructured spatial and temporal random effects interact with each other. This is a modification from what is referred to as Type I interaction in \cite{knorr2000bayesian}, also used by \cite{mercer2015space} in the spatio-temporal SAE literature. 
We chose the temporal variation to differ at the province level, as the district-level interaction model did not lead to a clear improvement (see Appendix \ref{appendix: sensitivity}).

We assign penalized complexity priors \citep{riebler2016intuitive,simpson2017penalising, sorbye2017penalised} on the scale parameters $\boldsymbol{\sigma}=(\sigma_\phi$, $\sigma_\zeta$, $\sigma_\nu$, $\sigma_\phi, \sigma_\psi)^\top$. We specify the priors so that $Pr(\sigma_u>1) = 0.01$ for all $\sigma_u\in\boldsymbol{\sigma}$. This weakly informative prior reflects the belief that the scale parameters are highly likely between 0 and 1. 
We conducted a sensitivity analysis by assigning different priors, such as the half-Cauchy distribution, and found little difference in both predictive performance and posterior densities of hyperparameters (see Appendix \ref{appendix: sensitivity}). 
For the regression coefficients $\boldsymbol{\beta}$, we place independent normal distributions with variance $5^2$.

We specified and fitted the model using the probabilistic programming language \texttt{Stan} \citep{Stan}, and we used the No-U-Turn sampler (NUTS) for inference. To implement the ICAR model in \texttt{Stan} we followed \cite{morris2019bayesian}.
We ran 4 chains for 1500 iterations with the first 500 iterations discarded as warm-up.

\subsection{Poststratification for subpopulation-level inference}\label{section: poststratification}
In the context of real-time small-area estimation, the subpopulation of interest is the intersection of geographical area (district) and time point (month). The first step of the poststratification involves partitioning the population into mutually exclusive sub-groups, referred to as “poststratification cells.” These cells are defined based on a set of categorical variables—typically related to demographic, geographical, and socio-economic status—called poststratification variables. We use $j$ to index these poststratification cells.

Each cell $j$ is defined by a unique combination of poststratification variables and represents a subgroup within the population. We denote the relative size of cell $j$ in the population by $n_j$.
Suppose we aim to estimate the mean of an outcome variable $Y$ for a specific subpopulation defined by a subset of cells, indexed by the set $\mathcal{J}$. We denote this subpopulation mean by $\theta^{(\mathcal{J})}$.
We estimate the cell-specific means of the response variable, denoted by $\hat{y}_j$. The post-stratified estimate of the mean for subpopulation $\mathcal{J}$, denoted by $\hat{\theta}^{(\mathcal{J})}$, is computed as a weighted average of the cell-level estimates: $\hat{\theta}^{(\mathcal{J})} = \frac{\sum_{j\in \mathcal{J}} n_j \hat{y}_j}{\sum_{j\in\mathcal{J}}n_j}$.
While a common choice for the cell-level estimate $\hat{y}_j$ is the sample mean within each cell, this approach can lead to issues when the number of cells is large, for example, when many poststratification variables or variables with a large number of categories are included. In such cases, some cells may contain few or no sampled units, making the estimates highly unstable or undefined. MRP addresses this issue by estimating the cell-specific means using a multilevel regression model, which borrows strength across cells to improve efficiency. 

\begin{table}[]
    \caption{Overview of different types of direct and modelled estimates used to calculate the prevalence of households with poor food consumption patterns. The table compares various estimates based on the type of regression model used, along with the aggregation method applied for subpopulation level estimates. Direct estimates, such as those from ZimVAC and mVAM, use design-based weights or sample proportions. Modelled estimates, including MR, MRP, and joint models (jMR), apply various aggregation methods, including simple aggregation of posterior predictive values and poststratification with cell weights obtained using the 2015 DHS, 2022 Census, and ZimVAC 2023 survey.}
    \centering
    \begin{tabular}{p{2.85cm}p{2.65cm}p{8.5cm}}
    \hline
    Type of estimates & Regression model  & Aggregation for subpopulation level estimate                                                                                                                  \\ \hline
    ZimVAC direct     & -                 & Sample proportion, see \eqref{eq: zimvac direct estimates}.                                                                                                                                             \\
    mVAM direct       & -                 & Weighted aggregation with design-based weights, given by \eqref{eq: mVAM direct estimates}.                                                                                                                \\
    MR                & Mobile-phone only & Simple aggregation of posterior predictive values in \eqref{eq: mobile-phone only MR estimates}.                                                                                                            \\
    MRP (DHS-Census)  & Mobile-phone only & poststratification of posterior predictive values, with cell weights estimated by raking 2015 DHS to match 2022 census margins. See \eqref{eq: mobile-phone only MRP}.                                \\
    jMR-MP            & Joint             & Simple aggregation of posterior predictive values, with survey modality set to mobile phone. See \eqref{eq: mobile-phone only MR estimates} and \eqref{eq: yhat jMR-MP}.                                                                    \\
    jMR-F2F           & Joint             & Simple aggregation of posterior predictive values, with survey modality set to F2F. See \eqref{eq: mobile-phone only MR estimates} and \eqref{eq: yhat jMR-F2F}.                                                                             \\
    jMRP (DHS-census) & Joint             & poststratification of posterior predictive values, with survey modality set to F2F \eqref{eq: joint MRP} and cell weights estimated by raking 2015 DHS to match 2022 census margins. \\
    jMRP (ZimVAC)     & Joint             & poststratification of posterior predictive values, with survey modality set to F2F \eqref{eq: joint MRP} and cell weights given by the frequency counts of ZimVAC 2023 survey.\\ \hline
    \end{tabular}
    \label{tab: types of estimates}
\end{table}

The target quantity in our analysis is the prevalence of households with poor food consumption patterns in district $s$ at time $t$, denoted by $p_{st}$. The poststratification variables include \textit{district} and covariates 2–7 from Table \ref{table:covariates}, namely \textit{Water source, Head of household education, Female head of household, Household size, Household toilet type}, and \textit{Household mobile phone ownership}. After fitting our model \eqref{eq: main logit model} jointly to the mVAM and ZimVAC datasets, we obtain posterior predictive distributions of the response variable $Y$ for every combination of the poststratification variables—i.e., for each cell—with the survey mode fixed to F2F.
We assume that the joint distribution of poststratification variables, or equivalently, the relative size of each cell within a district, does not change over the study period. We denote the index set for district $s$ by $\mathcal{J}_s$. The joint MRP (jMRP) estimate for $p_{st}$ is given by:
\begin{equation}\label{eq: joint MRP}
\hat{p}_{st}^\text{jMRP} = \frac{\sum_{j\in\mathcal{J}_s} n_j \hat{y}_{j,st}^{\mathcal{M}_\text{Joint}}}{\sum_{j\in\mathcal{J}_s}n_j}
\end{equation}
where
\begin{equation}\label{eq: cell estimates jMRP}
    \hat{y}_{j,st}^{\mathcal{M}_\text{Joint}} = \mathbb{E}_{\mathcal{M}_\text{Joint}}\left[Y | x_{2-7,j}, \text{Modality} = \text{F2F}, \text{District} = s, \text{Month} = t, \mathcal{D}\right]
\end{equation}
is the posterior mean of $Y$ under the multilevel regression model $\mathcal{M}_\text{Joint}$ and the observed data $\mathcal{D}=\{y_i,\mathbf{x}_i\}_{i=1}^N$. The key assumption here is that the F2F survey adequately represents the target population, including both households with and without access to a mobile phone, and that the cell-specific estimate \eqref{eq: cell estimates jMRP} is unbiased for the corresponding cell-specific population mean.
In addition to adjusting for sampling bias by including poststratification variables $x_{2\text{–}7}$ in the regression model, including survey modality as a covariate serves two further purposes. First, it helps mitigate the risk of residual sampling bias due to unmeasured or omitted variables that may be associated with both survey participation and the outcome, as highlighted by \cite{gellar2023calibrated}. Second, it accounts for survey modality bias—that is, systematic differences in how respondents answer the same questions depending on whether the survey is conducted F2F or by phone. By fixing the survey mode to F2F when generating predictions, we aim to standardise the response variable as if data had been collected through face-to-face interviews.

In practice, a fully Bayesian approach to obtaining the jMRP estimate of $p_{st}$, the prevalence in district $s$, within province $r$, at time $t$, and its posterior distribution involves the following steps. Given the model $\mathcal{M}_\text{Joint}$, data $\mathcal{D}$, and a set of priors on the parameters $\gamma, \boldsymbol{\beta}, \phi_s, \zeta_s, \nu_t, \xi_t, \psi_{rt}$ and hyperparameters $\boldsymbol{\sigma}$:

\begin{enumerate}
    \item Draw a sample of size $B$ from the joint posterior distribution of the parameters. We denoted this sample by $\{\gamma^{(b)}, \boldsymbol{\beta}^{(b)}, \phi_s^{(b)}, \zeta_s^{(b)}, \nu_t^{(b)}, \xi_t^{(b)}, \psi_{rt}^{(b)}\}_{b=1}^B.$
    \item For each $b = 1, \ldots, B$, and for each poststratification cell $j \in \mathcal{J}_s$ with covariate vector $\mathbf{x}_j$, draw $y_{j,st}^{(b)}$ from a Bernoulli distribution with success probability $\pi_{j,st}^{(b)}$, where 
    \begin{equation*}
    \pi_{j,st}^{(b)} = \frac{1}{1 + \exp\left(-\eta_{j,st}^{(b)}\right)}, \quad \eta_{j,st}^{(b)} = \gamma^{(b)} + \mathbf{x}_j^\top \boldsymbol{\beta}^{(b)} + \phi_s^{(b)} + \zeta_s^{(b)} + \nu_t^{(b)} + \xi_t^{(b)} + \psi_{rt}^{(b)}.
    \end{equation*}
    \item Compute $\hat{p}_{st}^{(b)}$ using \eqref{eq: joint MRP}, replacing $\hat{y}^{\mathcal{M}_\text{Joint}}_{j,st}$ with $y_{j,st}^{(b)}$ for each posterior draw $b = 1, \ldots, B$.
\end{enumerate}

\noindent The resulting sample $\hat{p}_{st}^{(1)}, \ldots, \hat{p}_{st}^{(B)}$ can then be used to compute posterior summaries of interest, including the posterior mean, quantiles, and credible intervals. Note that we apply this procedure to obtain all model-based estimates (see Table~\ref{tab: types of estimates}) presented in this paper.

Thus far, we have assumed that the relative cell sizes in the population, or poststratification weights $n_j$ are known; however, this is rarely the case in practice. When a survey that is representative at the desired level (e.g., district) is available, as with the F2F ZimVAC survey in our application, frequency counts for each poststratification cell can be directly obtained and used as cell weights. In the absence of such data, cell weights must be estimated. A common approach in this setting is iterative proportional fitting (IPF), or raking \citep{deming1940least, deville1992calibration, deville1993generalized}, which adjusts sample weights to match known population marginals. The procedure generates adjusted weights $w_j$ that minimise the distance to a set of prior weights (e.g., design weights $d_j$), while ensuring the weighted marginals align with target population distributions. For mobile-phone-only models, where we assume no recent district-level representative survey is available, we apply this approach by using the 2015 DHS data (representative at the province level) as prior weights, and rake them to match marginal distributions from the 2022 Census.

For the mobile-phone-only model \(\mathcal{M}_\text{MPonly}\), we follow the same procedure to obtain jMRP estimates as in \eqref{eq: joint MRP}. However, since we do not include a variable to adjust for modality bias, we replace the cell-specific estimates in \eqref{eq: cell estimates jMRP} with  
$
\hat{y}_{j,st}^{\mathcal{M}_\text{MPonly}} = \mathbb{E}_{\mathcal{M}_\text{MPonly}}[Y \mid x_{2\text{--}7,j}, \text{District} = s, \text{Month} = t,\mathcal{D}],
$
and compute the MRP estimate for prevalence as  
\begin{equation}\label{eq: mobile-phone only MRP}
\hat{p}_{st}^\text{MRP} = \frac{\sum_{j \in \mathcal{J}_s} w_j \hat{y}_{j,st}^{\mathcal{M}_\text{MPonly}}}{\sum_{j \in \mathcal{J}_s} w_j}.
\end{equation}

\subsection{Validation and comparison}\label{section: methods validation and comparison}
To assess the performance of our jMRP and MRP estimates, we validate them against direct estimates from the Rural ZimVAC survey, defined as the sample proportion of households with poor food consumption patterns in district $s$ at time $t$:
\begin{equation}\label{eq: zimvac direct estimates}
        \hat{p}_{st}^\text{ZimVAC} = \frac{\sum_{i\in \mathcal{I}_{st}^\text{ZimVAC}} y_i}{n_{st}^\text{ZimVAC}}.
\end{equation}
 Here, $n_{st}^\text{ZimVAC}$ denotes the number of households in district $s$ at time $t$ in the ZimVAC survey, and $\mathcal{I}_{st}^\text{ZimVAC}$ is the set of indices corresponding to this subgroup, such that $|\mathcal{I}_{st}^\text{ZimVAC}|=n_{st}^\text{ZimVAC}$. We treat the ZimVAC direct estimates as ground truth and evaluate our modelled estimates in terms of both predictive accuracy and uncertainty quantification. As our analysis uses the annual Rural ZimVAC surveys conducted in 2023 and 2024, we validate the estimates at two time points.

 To assess the accuracy of our modelled estimates, we compute a range of metrics including Mean Absolute Error (MAE), Root Mean Squared Error (RMSE), Mean Bias Error (MBE), Pearson’s and Spearman’s correlation coefficients, and Lin’s Concordance Correlation Coefficient (CCC) \citep{lawrence1989concordance}. CCC evaluates the degree to which pairs of observations fall on the 45-degree line through the origin, thereby assessing both precision and accuracy in agreement between the modelled and direct estimates.

To evaluate uncertainty quantification, we examine coverage and interval length of the credible intervals, as well as the Continuous Ranked Probability Score (CRPS). Here, coverage is computed as the proportion of instances in which the credible interval of the modelled estimate overlaps with the confidence interval of the corresponding ZimVAC estimate. For the ZimVAC direct estimates, we use the Wilson score interval rather than the Wald interval; the rationale for this choice is discussed below.

To further evaluate the performance of our jMRP and MRP estimates, we compare them with alternative estimates: (i) direct estimates from the mVAM survey, and (ii) multilevel regression (MR) estimates without poststratification. The latter are obtained by fitting the same spatio-temporal model described in Section~\ref{section: model}, but aggregating predictions at the unit level, without adjusting for population structure. Table~\ref{tab: types of estimates} summarises all direct and modelled estimates considered in our analysis. 
The mVAM direct estimate of $p_{st}$ is computed as:
 \begin{equation}\label{eq: mVAM direct estimates}
        \hat{p}_{st}^\text{mVAM} = \frac{\sum_{i\in \mathcal{I}^\text{mVAM}_{st}} w^\text{mVAM}_i y_i}{\sum_{i\in \mathcal{I}^\text{mVAM}_{st}}w^\text{mVAM}_i}.
\end{equation}
 where $\mathcal{I}^\text{mVAM}_{st}$ is the set of observations in district $s$ at time $t$, and $w^\text{mVAM}_i$ denotes the weight associated with observation $y_i$. Details on the construction of these weights are provided in Appendix~\ref{appendix: mVAM data processing}. We compute confidence intervals for the mVAM estimates using the Wilson score interval. This choice accounts for the fact that the mVAM survey is designed for province-level analysis, leading to small sample sizes at the district level. In such cases, extreme prevalence values (0\% or 100\%) are common, and the standard Wald interval can yield confidence intervals of length zero. While the ZimVAC survey does not face the same issue, we use the Wilson interval for consistency. 
 
For the mobile-phone-only model, the MR estimates are computed by averaging the posterior mean from the fitted regression model across individual observations, without applying poststratification
 \begin{equation}\label{eq: mobile-phone only MR estimates}
        \hat{p}_{st}^\text{MR} = \frac{\sum_{i\in \mathcal{I}^\text{mVAM}_{st}} \hat{y}_i^{\mathcal{M}_\text{MPonly}}}{n_{st}^\text{mVAM}}.
\end{equation}
where $\hat{y}^{\mathcal{M}_\text{MPonly}}_i = \mathbb{E}_{\mathcal{M}_\text{MPonly}}\left[Y | x_{2-7,i}, \text{district} = s[i], \text{month} = t[i],\mathcal{D} \right]$.
"These estimates are not conventional unit-level SAE estimates, but are included to assess the contribution of the poststratification step. This is known as an ablation study in the machine learning literature.s

For the joint model, we compute two variants of MR estimates by fixing the survey modality to either mobile phone (MP) or face-to-face (F2F). These joint MR (jMR) estimates are given by: 
\begin{align}
 \hat{y}_i^{\mathcal{M}_\text{joint},\text{MP}} &= \mathbb{E}_{\mathcal{M}_\text{Joint}}\left[Y | x_{2-7,i}, \text{Modality}=\text{MP},\text{District} = s[i], \text{Month} = t[i] \right] \quad \text{and} \label{eq: yhat jMR-MP} \\
 \hat{y}_i^{\mathcal{M}_\text{Joint}, \text{F2F}} &= \mathbb{E}_{\mathcal{M}_\text{Joint}}\left[Y | x_{2-7,i}, \text{Modality}=\text{F2F}, \text{District} = s[i], \text{Month} = t[i] \right] \label{eq: yhat jMR-F2F} 
\end{align}
and the resulting averages, computed using the same approach as in \eqref{eq: mobile-phone only MR estimates}, yield the jMR-MP and jMR-F2F estimates, respectively.

We evaluate model performance at two time points using the Rural ZimVAC surveys conducted in May 2023 and 2024. For the joint model, the 2023 ZimVAC data is included in model training; thus, any evaluation of the corresponding modelled estimates at that time point reflects within-sample predictive performance. In contrast, the 2024 ZimVAC data is reserved exclusively for evaluation, allowing us to assess out-of-sample performance. For the mobile-phone-only model, the ZimVAC survey is not used in model fitting, so performance at both time points reflects out-of-sample prediction.

\subsection{Real-time monitoring and nowcasting}
Beyond retrospective calibration, our jMRP framework is well-suited for real-time monitoring and nowcasting of food insecurity. By holding out the 2024 ZimVAC data and training the model using only the 2023 survey, we simulate a scenario in which past surveys—conducted in a preferred modality and representative of the target population at the desired level—are used to calibrate more frequently collected, up-to-date surveys that differ in modality and are designed for analysis at higher administrative levels. The temporal random effects in the multilevel regression capture dynamic trends informed by high-frequency data, while including survey modality as a covariate helps mitigate response bias. poststratification further reduces sampling bias and variance, enabling the production of timely, accurate, and well-calibrated estimates.

\section{Results} \label{section: result}
Our approach begins by fitting the multilevel regression model~\eqref{eq: main logit model} at the household level, with the results presented in Section~\ref{section: covariates effect}. Cell-level predictions from the model are subsequently aggregated to district-level monthly estimates via poststratification. The accuracy and uncertainty of the jMRP and MRP estimates are evaluated in Section~\ref{section: result validation and comparison}, while Section~\ref{section: result RTM district level} examines the spatial and temporal variations in the estimated prevalence of households with poor food consumption patterns. 

\subsection{Covariate effects}\label{section: covariates effect}
Table~\ref{Table: regression parameter estimates} presents the parameter and hyperparameter estimates for two models: the mobile-phone-only model and the joint model. As expected, when the modality is set to mobile phone, the joint model’s results closely align with those of the mobile-phone-only model. Both models indicate that households headed by females are associated with higher odds of poor food consumption. In contrast, improved living conditions—such as access to improved water sources and toilets—along with mobile phone ownership and higher education levels, are linked to lower odds of poor food consumption.

\begin{table}[t]
\centering
\caption{Posterior estimates of covariate effects and of
parameters associated with the spatio-temporal effects}
\label{Table: regression parameter estimates}
\begin{tabular}{lrcrc}
\hline
\multicolumn{1}{c}{Model} & \multicolumn{2}{c}{Mobile-phone only} & \multicolumn{2}{c}{Joint} \\
                          & \multicolumn{1}{c}{Mean} & \multicolumn{1}{c}{95\% CI} & \multicolumn{1}{c}{Mean}  & \multicolumn{1}{c}{95\% CI} \\ [0.5ex]  \hline
\textbf{Covariates}       & & \\
0: Intercept & -1.453 & (-1.775 , -1.134) & -1.375 & (-1.573,-1.186) \\
1: Modality of survey (F2F)  &  &  & 1.438&  (\textcolor{white}{- }1.138,\textcolor{white}{-}1.755) \\
2: Water source (improved) &-0.489 & (-0.557, -0.422) &-0.495 & (-0.565, -0.424)\\
3.1: Head of household education (primary)  & -0.126 & (-0.235,-0.019) & -0.101&( -0.200, -0.008)\\
3.2:  Head of household education (secondary)& -0.540 & (-0.644, -0.437)  & -0.509& (-0.581, -0.433) \\
3.3: Head of household education (higher)  & -1.692 & (-1.935, -1.466) & -1.653 &(-1.866, -1.444) \\
  4: Female head of household      & 0.289 & (\textcolor{white}{-}0.227, \textcolor{white}{-}0.351) & 0.289 & (\textcolor{white}{-}0.227, \textcolor{white}{-}0.349) \\
5.1: Household size (3-4 people) & 0.187 & (\text{ }0.018,\text{ }0.370) & 0.192 &  (\text{ }0.016,\text{ }0.364)\\
5.2: Household size (5-6 people) & 0.463 & (\text{ }0.295,\text{ }0.638) & 0.474 & (\text{ }0.306,\text{ }0.639) \\
5.3: Household size (7+ people)  & 0.660 & (\text{ }0.476,\text{ }0.841)& 0.681 &(\text{ }0.510,\text{ }0.848)\\
  6: Household toilet type (improved) &-0.198 & (-0.266, -0.132) & -0.194& (-0.257, -0.133) \\
  7: Household mobile phone ownership &-0.366 & (-0.885, \text{ }0.116)& -0.557 & (-0.649, -0.465) \\ \hline
  8: 1(F2F) $\times$ 2 (Improved Water Source) & & & 0.403& (\textcolor{white}{-}0.284, \textcolor{white}{-}0.526) \\
9.1: 1(F2F) $\times$ 3.1 (education=primary) & & & -0.014& (-0.177, \textcolor{white}{-}0.152) \\
9.2: 1(F2F) $\times$ 3.2 (education=secondary) & & & 0.259& (\textcolor{white}{-}0.104,  \textcolor{white}{-}0.421)\\
9.3: 1(F2F) $\times$ 3.3 (education=higher) & & & -0.157& (-0.849, \textcolor{white}{-}0.482)\\
 10: 1(F2F) $\times$ 4 (Female head of household) & & & -0.110& (-0.223,  0.004)\\
11.1: 1(F2F) $\times$ 5.1 (household size = 3-4) & & &-0.302& (-0.507, -0.085)\\
11.2: 1(F2F) $\times$ 5.2 (household size = 5-6) & & &-0.584& (-0.791, -0.373)\\
11.3: 1(F2F) $\times$ 5.3 (household size = 7+) & & &-0.829& (-1.55, -0.603)\\
 12: 1(F2F) $\times$ 6 (Improved toilet type) & & &-0.262& (-0.371, -0.157)\\ 
 [0.5ex]  \hline
\textbf{Hyper-parameters} & & \\
$\sigma_\phi$  &  0.463 & (0.262, 0.639)  &  0.515 &  (0.235, 0.770) \\
$\sigma_\zeta$& 0.112 & (0.005, 0.251)  & 0.202 &  (0.022, 0.344) \\
$\sigma_\nu$  & 0.197 & (0.131, 0.271)&  0.197&  (0.137,  0.272) \\
$\sigma_\xi$  & 0.056 & (0.002, 0.153)& 0.051&  (0.002, 0.142) \\
$\sigma_\psi$  &0.085 & (0.006, 0.158)  & 0.162 &  (0.125,  0.203)\\
\hline
\end{tabular}
\end{table}

The joint model offers further insights when considering the effects of covariates by survey modality. For example, the joint model shows that the impact of a female head of household is stronger in the mobile-phone survey, with the odds of poor food consumption being 33.5\% higher compared to male-headed households. In the F2F survey, this increase is smaller, at 19.6\%. This illustrates how data collection modality influences the relationship between household characteristics and food consumption patterns. A similar difference is observed in the effect of access to improved water sources. In the mobile-phone survey, access to improved water reduces the odds of poor food consumption by 39.0\%, while the F2F survey shows a much smaller effect (5.5\% decrease). 

Notably, when comparing two households with identical characteristics, the one surveyed via F2F is more likely to exhibit poor food consumption patterns than the one surveyed by mobile phone. This finding reinforces the presence of modality bias, suggesting that survey mode can influence responses and introduce systematic bias into cell-level estimates, which in turn affects subpopulation-level estimates.

\subsection{Validating and comparing the estimates}\label{section: result validation and comparison}
Tables~\ref{tab: validation correlation and sharpness} and~\ref{tab: uncertainty} summarise the performance of different models and methods for estimating district-level prevalence of poor food consumption. The evaluation focuses on a range of predictive performance metrics, including correlation (Pearson, Spearman, and CCC), error (RMSE, MAE, and bias), coverage, and uncertainty quantification (e.g., CRPS and confidence interval lengths). Performance is assessed both in-sample (2023) and out-of-sample (2024).

As discussed, the mVAM survey in Zimbabwe is designed for province-level estimation, and thus sampling is not conducted with district-level representativeness in mind. Direct mVAM estimates, which involve no modelling, exhibit high variability due to small sample sizes at the district level. This is particularly evident in some districts where all surveyed households fall into the poor food consumption group. This highlights the challenge of deriving reliable district-level estimates from a survey designed for higher administrative levels, underscoring the need for modelling approaches to produce more robust and consistent district-level estimates.

\begin{figure}
    \centering
    \includegraphics[width=0.95\linewidth]{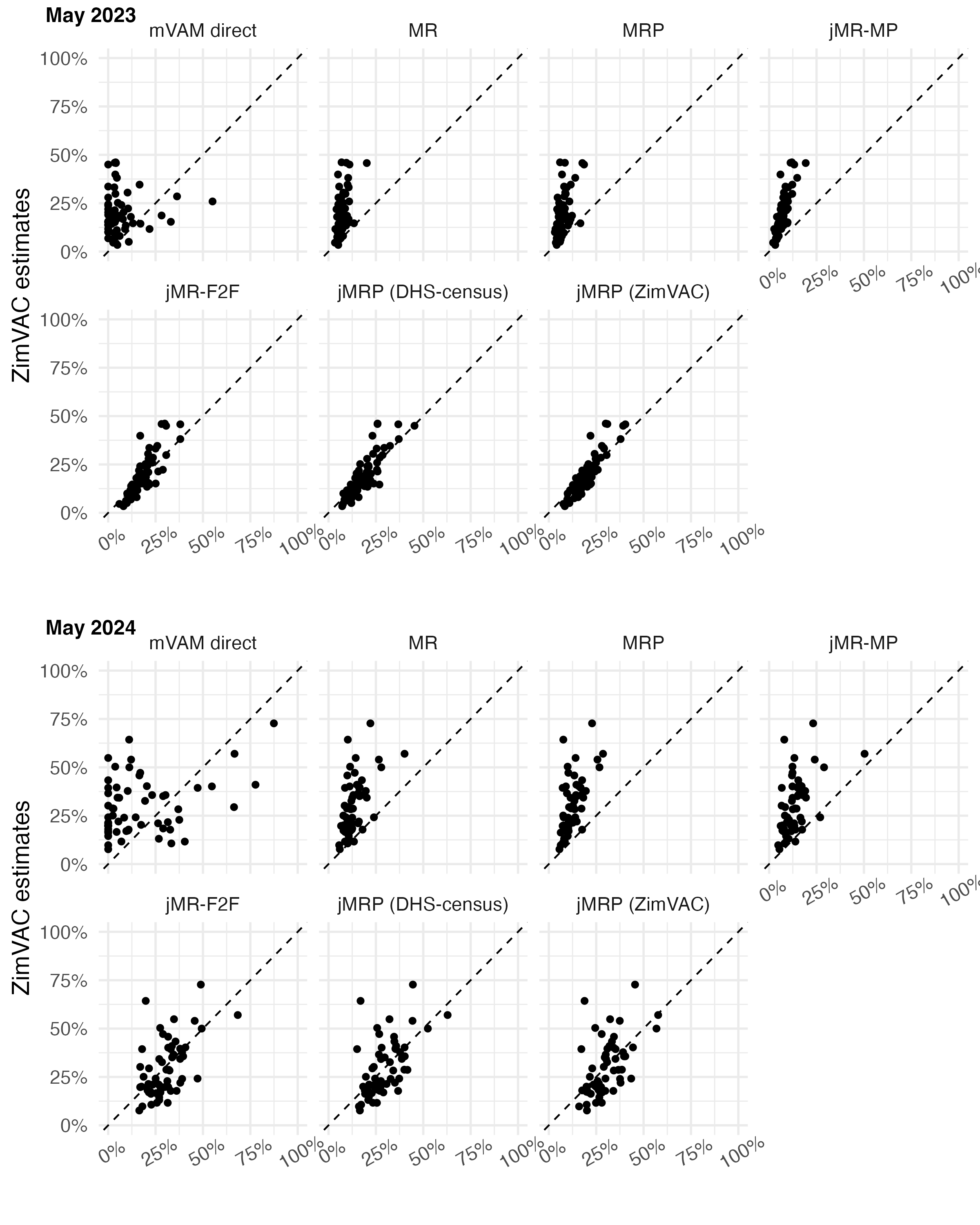}
    \caption{Scatter plots comparing modelled estimates to ZimVAC ground truth estimates for May 2023 and May 2024.
This figure evaluates the performance of different models by comparing their poor food consumption prevalence estimates to the ZimVAC ground truth estimates. Panel A presents the comparison for May 2023, while Panel B shows the comparison for May 2024. The dashed line $y=x$ indicates perfect agreement between the modelled estimates and the ground truth, with points closer to the line reflecting better model performance.}
    \label{figure: correlation}
\end{figure}

Modelled estimates based on mobile-phone-only data, including both MR and MRP (DHS-Census), show incremental improvements in correlation and error metrics, with MRP generally performing better, for example, achieving a Pearson correlation of 0.632 in May 2024 compared to 0.574 for MR. However, both models exhibit consistent negative bias, indicating underestimation of poor food consumption prevalence (e.g., MBE = –0.161 for MR and –0.172 for MRP in May 2024). This is also evident in the scatter plots in Figure~\ref{figure: correlation}, where mobile-phone-only estimates fall below the diagonal line.

\begin{figure}[t!]
\centering
\includegraphics[width=0.95\linewidth]{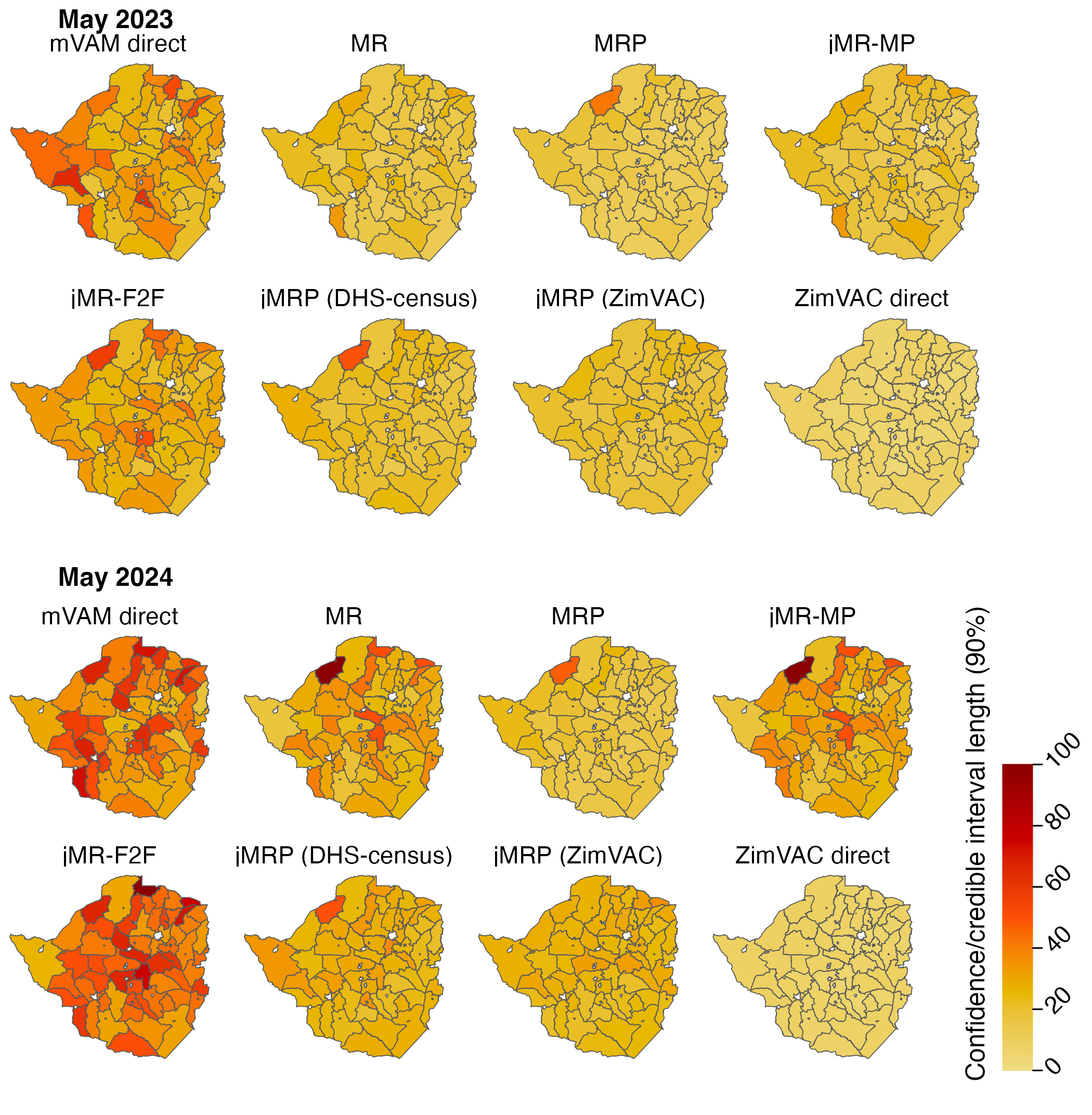}
\caption{Comparison of the 90\% confidence/credible interval length for estimates of prevalence of households with poor food consumption patterns across different models and methods. The figure compares the interval lengths for both direct estimates (mVAM and ZimVAC, confidence interval) and modelled estimates (MR, MRP, jMR-MP, two types of jMR, and jMRP, credible interval) for May 2023 (top panel) and May 2024 (bottom panel). The colour scale ranges from yellow (0) to red (100), with higher values indicating greater uncertainty in the estimates.}
\label{figure: map ci length}
\end{figure}

We now turn to the evaluation of the joint model-based estimates. Comparison between the MR estimates with survey modality fixed to either mobile phone or F2F highlights the value of adjusting for modality bias. For instance, in May 2023, the joint MR model with F2F modality (jMR-F2F) achieves a much higher CCC (0.740) and lower mean bias (–0.012) than its mobile-phone counterpart (jMR-MP: CCC = 0.296, MBE = –0.127). However, without poststratification, the jMR-F2F estimates yield more conservative uncertainty quantification, as reflected in wider confidence intervals and higher CRPS values (e.g., for jMR-F2F: 90\% CI length = 0.477, CRPS = 0.069), compared to the post-stratified joint MRP estimates (e.g., jMRP (ZimVAC): 90\% CI length = 0.262, CRPS = 0.066; see Figure~\ref{figure: map ci length}).

For jMRP estimates, poststratification cells and their weights are derived either by raking DHS data to match Census marginals or directly from the ZimVAC 2023 survey (see Table~\ref{tab: types of estimates}). In addition to achieving narrower credible intervals while maintaining coverage close to nominal levels, jMRP estimates also exhibit lower bias. Both jMRP (DHS-Census) and jMRP (ZimVAC) perform well, with the latter achieving the strongest in-sample performance in 2023 (Pearson = 0.901, rMSE = 0.053, bias $<$ 0.001), and the former excelling out-of-sample in 2024 (Pearson = 0.568, rMSE = 0.118, bias = –0.001). These models demonstrate strong agreement with the ZimVAC direct estimates, as illustrated by high correlation and CCC values (above 0.8 in-sample, around 0.5 out-of-sample), and by the tighter clustering of points along the identity line in Figure~\ref{figure: correlation}, indicating close alignment with the reference estimates.

\begin{table}[t]
\caption{Performance metrics for different models and methods estimating district-level poor food consumption prevalence, evaluated for in-sample (2023) and out-of-sample (2024) settings. Metrics include Pearson and Spearman correlation coefficients, concordance correlation coefficient (CCC), root mean square error (rMSE), mean absolute error (MAE), and mean bias error (MBE). For details on the types of estimates, refer to Table \ref{tab: types of estimates}.}
\label{tab: validation correlation and sharpness}
\centering
\begin{tabular}{llrrrrrr}
\hline
Month    & Type of estimates& Pearson & Spearman & CCC   & rMSE  & MAE   & MBE             \\ \hline
2023 May & mVAM direct      & 0.051 & 0.073 & 0.028 & 0.194 & 0.159 & -0.131\\
         & MR               & 0.484 & 0.417 & 0.101 & 0.159 & 0.127 & -0.126\\
         & MRP (DHS-Census) & 0.509 & 0.479 & 0.131 & 0.155 & 0.124 & -0.123\\
         & jMR-MP           & 0.823 & 0.810 & 0.196 & 0.152 & 0.127 & -0.127\\
         & jMR-F2F          & 0.829 & 0.838 & 0.740 & 0.065 & 0.047 & -0.012\\ 
         & jMRP (DHS-Census)& 0.831 & 0.827 & 0.782 & 0.062 & 0.044 & -0.004\\
         & jMRP (ZimVAC)    & \textbf{0.901}   & \textbf{0.888}    & \textbf{0.836} & \textbf{0.053} & \textbf{0.039} & $<$\textbf{0.001} \\ \hline
2024 May & mVAM direct      & 0.333 & 0.244 & 0.248 & 0.245 & 0.210 & -0.127\\
         & MR               & 0.562 & 0.535 & 0.186 & 0.201 & 0.162 & -0.161\\
         & MRP (DHS-Census) & \textbf{0.632} & \textbf{0.605} & 0.173 & 0.208 & 0.172 & -0.172\\
         & jMR-MP           & 0.521 & 0.536 & 0.201 & 0.202 & 0.163 & -0.161\\
         & jMR-F2F          & 0.539 & 0.527 & 0.495 & 0.122 & 0.093 & 0.008\\
         & jMRP (DHS-Census)& 0.568 & 0.584 & \textbf{0.519} & \textbf{0.118} & \textbf{0.087} & \textbf{-0.001}\\
         & jMRP (ZimVAC)   & 0.534 & 0.549 & 0.474 & 0.121 & 0.091 & 0.005\\
         \hline
\end{tabular}
\end{table}

\begin{table}[t]
    \caption{Uncertainty quantification metrics for different types of estimates of district-level poor food consumption prevalence. For details on the types of estimates, refer to Table \ref{tab: types of estimates}. Metrics include coverage and confidence interval (CI) length for 80\% and 90\% intervals, as well as continuous ranked probability score (CRPS), evaluated for in-sample (2023) and out-of-sample (2024) settings.}
    \label{tab: uncertainty}
    \centering
    \begin{tabular}{llrrrrr}
    \hline
    Month    & Type of Estimates      & \multicolumn{2}{c}{80\% CI} & \multicolumn{2}{c}{90\% CI} & CRPS  \\ 
             &                       & Coverage      & Length      & Coverage      & Length      &       \\ \hline
    2023 May & mVAM direct            & 0.717         & 0.241       & 0.900         & 0.321       &-      \\
             & MR                     & 0.550         & 0.156       & 0.683         & 0.189       & 0.108 \\
             & MRP (DHS-Census)       & 0.383         & 0.108       & 0.583         & 0.138       & 0.106 \\
             & jMR-MP                 & 0.500         & 0.156       & 0.700         & 0.182       & 0.105 \\
             & jMR-F2F                & 0.950         & 0.242       & 0.983         & 0.303       & 0.041 \\
             & jMRP (DHS-Census)      & 0.933         & 0.167       & 0.967         & 0.213       & 0.034 \\
             & jMRP (ZimVAC)          & 0.950         & 0.152       & 0.983         & 0.195       & \textbf{0.028} \\ \hline
    2024 May & mVAM direct            & 0.700         & 0.338       & 0.900         & 0.428       &-      \\
             & MR                     & 0.500         & 0.252       & 0.683         & 0.318       & 0.130 \\
             & MRP (DHS-Census)       & 0.317         & 0.139       & 0.467         & 0.178       & 0.145 \\
             & jMR-MP                 & 0.517         & 0.244       & 0.700         & 0.317       & 0.131 \\
             & jMR-F2F                & 0.950         & 0.369       & 0.983         & 0.477       & 0.069 \\
             & jMRP (DHS-Census)      & 0.867         & 0.218       & 0.900         & 0.277       & \textbf{0.063}\\
             & jMRP (ZimVAC)          & 0.767         & 0.206       & 0.917         & 0.262       & 0.066 \\ \hline
    \end{tabular}
\end{table}

\begin{figure}[t!]
\centering
\includegraphics[width=1.0\linewidth]{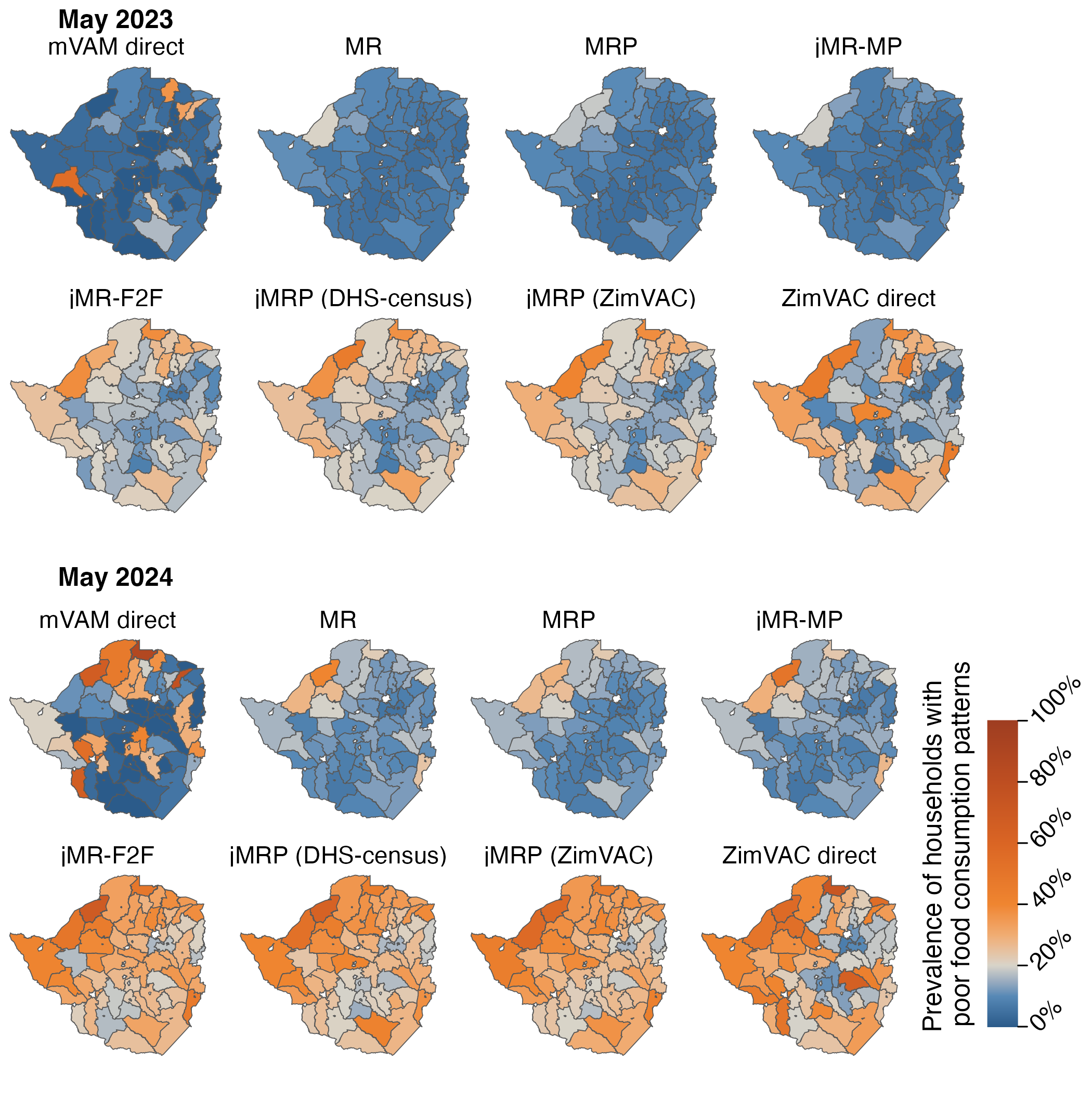}
\caption{Comparison of direct and modelled estimates for the prevalence of households experiencing poor food consumption patterns in rural districts of Zimbabwe for May 2023 (top panel) and May 2024 (bottom panel). Estimates are derived using various models and methods, with colours indicating the prevalence: blue for lower prevalence (0\%) and orange for higher prevalence (up to 100\%). The maps highlight differences across direct estimates (mVAM and ZimVAC) and modelled estimates, including MR, MRP, and jMRP variations.}
\label{figure: map validation}
\end{figure}

Figure~\ref{figure: map validation} presents the modelled and direct estimates of the prevalence of households with poor food consumption across rural districts. Compared to the direct mVAM estimates, the modelled estimates exhibit smoother spatial patterns. However, mobile-phone-only model estimates (MR and MRP) and the joint MR estimates with mobile-phone modality (jMR-MP) tend to underestimate the severity of poor food consumption, consistent with the findings discussed above. In contrast, the jMR-F2F and both jMRP estimates align more closely with the ZimVAC direct estimates, capturing similar spatial trends and identifying key hotspots in the northwestern regions, such as Kariba and Binga (see Figure~\ref{figure: zimbabewe map} for district locations), though some deviations remain—for example, in Gutu, Masvingo. As aforementioned, the credible intervals for two jMRP estimates are notably narrower than those for jMR-F2F, a difference clearly visible in Figure~\ref{figure: map ci length}. Additional results and validation analyses are provided in Appendix~\ref{appendix: additional results}.


\subsection{Visualising spatial and temporal pattern for real-time monitoring at district level}\label{section: result RTM district level}
\begin{figure}[t!]
    \centering
    \includegraphics[width=0.975\textwidth]{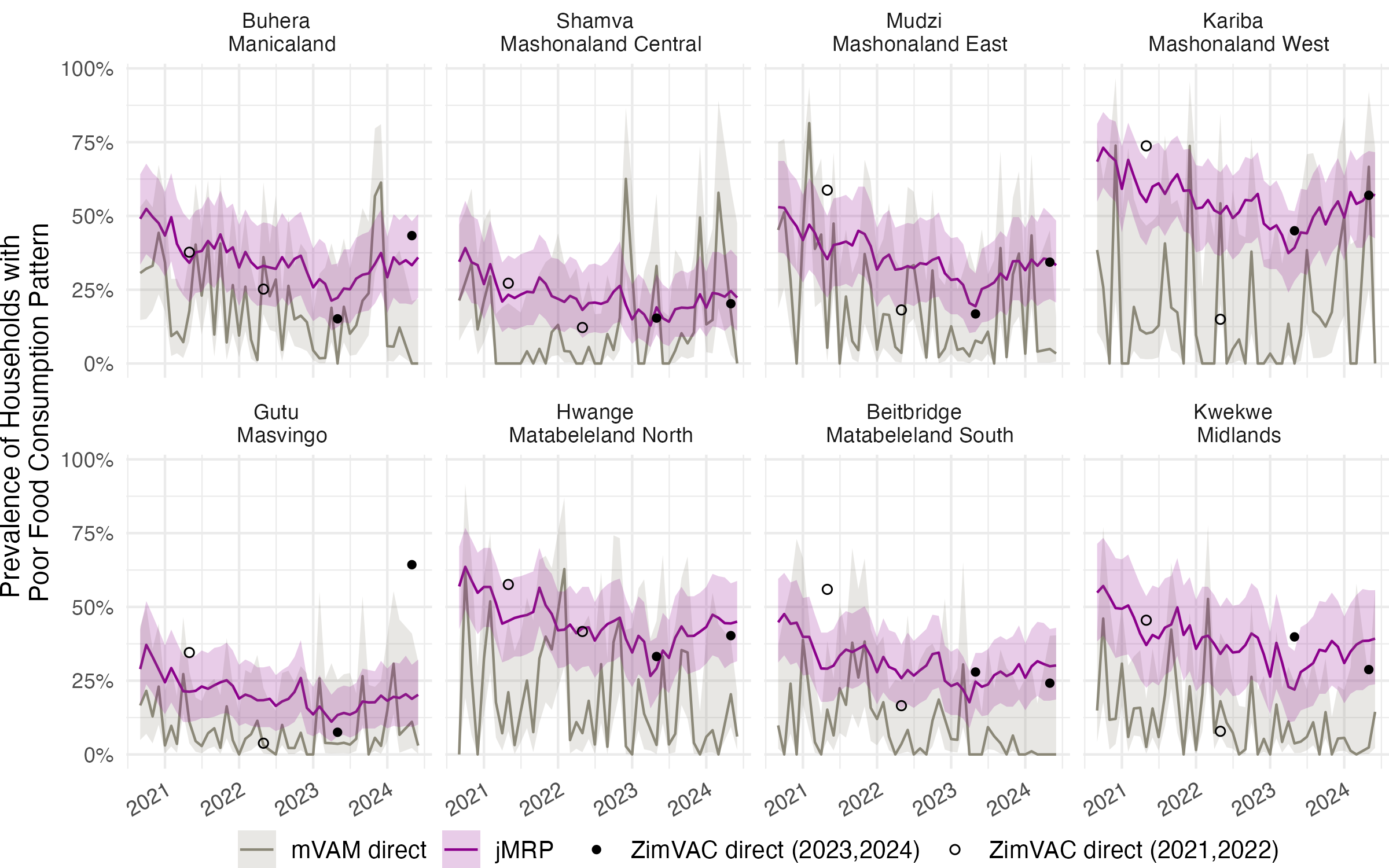}
    \caption{Prevalence of households with poor food consumption patterns across eight selected districts (September 2020 - June 2024). One district was chosen per province to represent both good and poor model fits. For time series plots of other districts, refer to Figure \ref{figure:time trend all 1} and Figure \ref{figure:time trend all 2}. The plots compare the prevalence of poor food consumption estimates and their uncertainty intervals: 90\% confidence intervals for direct mobile-phone-based (mVAM) estimates (grey), 90\% credible intervals for joint MRP (ZimVAC) estimates (magenta), and point estimates for ground truth ZimVAC data (black dots). ZimVAC estimates for 2021 and 2022 are provided for reference only, as the Food Consumption Score (FCS) was computed using a different set of food items.}
    \label{figure:time trend selected}
\end{figure}

From the right bottom panel (ZimVAC estimates) in Figure \ref{figure: map validation}, we see that the food security situation worsens from 2023 to 2024 in many districts in Zimbabwe. To understand the time trend of food security for different districts, we computed MRP estimates for the study period (September 2021 to June 2024) for all districts. Figure \ref{figure:time trend selected} illustrates the prevalence of households with poor food consumption patterns across multiple districts in Zimbabwe from September 2021 to June 2024. Compared to the mobile-phone-based mVAM estimates, the MRP estimates are much smoother and better align with the ground truth ZimVAC estimates in many districts. This is particularly evident across most regions where the MRP estimates follow the general trend of food consumption prevalence. Specifically, for the majority of districts, a high prevalence is observed in 2021, followed by a downward trend in 2022 and 2023, and then an increase again in 2024. The MRP model appears to capture these fluctuations in food consumption patterns with greater accuracy than the mVAM estimates, highlighting the improved predictive power of our approach.

However, regional variations are also apparent. While the MRP estimates closely match ZimVAC estimates in several districts, such as in Shamva and Hwange, there are areas, such as the Gutu district in the Masvingo province, where the MRP model struggles to capture sudden changes, particularly in 2024. This highlights the limitations of the MRP approach in capturing rapid, unforeseen shifts in food consumption patterns, which could be attributed to local events or policy changes that were not incorporated into the model. The estimated trends for all districts in Zimbabwe can be found in Figure \ref{figure:time trend all 1} and \ref{figure:time trend all 2} in Appendix \ref{appendix:FCS time series}.

\begin{figure}
    \centering
    \includegraphics[width=1.0\linewidth]{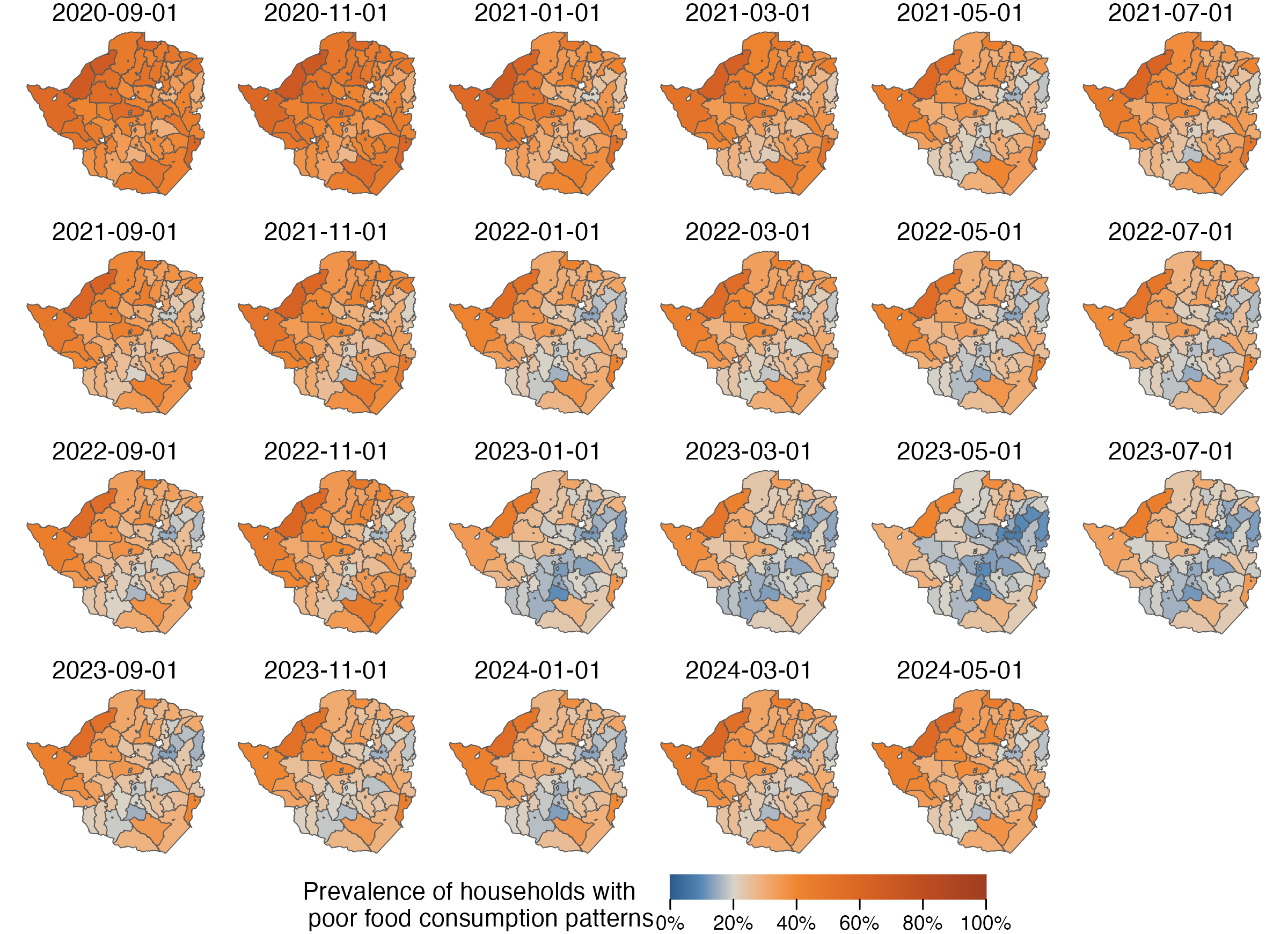}
    \caption{Estimated prevalence of households with poor food consumption patterns in Zimbabwe by month (September 2020 - June 2024). The series of maps shows the estimated prevalence of households with poor food consumption patterns in Zimbabwe from September 2020 to June 2024, as derived from the jMRP (ZimVAC) approach. Each map represents the monthly distribution of food security across the country, with varying levels of prevalence indicated by a colour scale from blue (0\%) to orange (100\%).}
    \label{fig: map prevalence over time}
\end{figure}

Similarly, the maps shown in Figure \ref{fig: map prevalence over time} highlight the dynamic and region-specific nature of food security. The figure presents the estimated prevalence of households with poor food consumption patterns across Zimbabwe from September 2020 to June 2024, based on the MRP (ZimVAC) model. At the beginning of the study period, in September 2020, the maps show high levels of poor food consumption patterns across many regions of the country, with large areas colored in orange and red, indicating a high prevalence of poor food consumption patterns. Over the course of the study period, there is a gradual improvement in food security, with significant reductions in the prevalence of poor food consumption, especially in central and southern regions, as reflected in the increase of green areas on the maps. However, from late 2023 onwards, the situation worsens again in certain regions, particularly in the north-west and south-east parts of Zimbabwe. These areas show a noticeable increase in poor food consumption prevalence, as seen in the rise of orange and red zones. This trend may reflect various factors such as seasonal fluctuations, economic challenges, or climatic events that could have impacted food availability and access, leading to a deterioration in food security for these regions.

It is important to note that while the MRP estimates generally provide useful insights, caution is needed when interpreting the results. In some districts, deviations from the ground truth (F2F survey direct estimates) have been observed, especially in areas where the model may not fully capture sudden shifts in food consumption patterns or regional-specific factors. These discrepancies underscore the limitations of the model and highlight the need for continuous validation against direct survey data to ensure the accuracy and reliability of the estimates.

\section{Summary and discussion}\label{section: conclusion}
This paper presented a methodology for real-time small area estimation of a food security indicator. Specifically, we introduced a jMRP approach that integrates high-frequency mobile phone survey data, designed for analysis at a higher administrative level, with an annual F2F survey that is representative at a lower administrative level. The two surveys complement each other: one provides frequent temporal updates, while the other offers finer spatial resolution. The multilevel regression step jointly models both data sources, accounting for modality bias and borrowing strength across time and space, while the poststratification step uses auxiliary variables, such as household head education and living conditions, to adjust for sampling bias and align estimates with the target population at the district level.

We applied our methodology to monitor the prevalence of poor food consumption in rural districts—the second administrative level—of Zimbabwe, a country regularly affected by severe climatic shocks. During the study period, food security was monitored through the mobile phone-based mVAM survey, providing frequent updates at the first administrative level, and the annual F2F Rural ZimVAC survey, representative at the second administrative level. The jMRP estimates were validated using ZimVAC direct estimates from May 2023 and May 2024. Compared to alternatives based solely on the mobile-phone survey or without poststratification, the jMRP approach substantially improved both accuracy and uncertainty quantification.  In out-of-sample (nowcasted) 2024 estimates, the best-performing jMRP estimates offer substantial improvements compared to mobile-phone-based direct estimates, reducing mean absolute error by 56.7\%, mean bias by 99.2\%, and producing credible intervals 38.8\% narrower while maintaining appropriate coverage. The jMRP approach also successfully captured deteriorating conditions across many districts, even without access to the most recent ZimVAC data, demonstrating its potential for real-time monitoring. 

The findings from this study provide a strong foundation and increased confidence to test and apply this methodology in other contexts. Specifically, they suggest that high-frequency mobile phone survey data, such as mVAM, can be efficiently utilised and calibrated to produce reliable small area estimates, even if originally designed for higher administrative levels. In doing so, the methodology addresses key sources of bias, including those related to survey modality and phone ownership. Importantly, this is achieved in a highly cost-efficient manner. While the level of precision may not match what could theoretically be achieved through large-scale F2F surveys conducted frequently and at high spatial granularity, such surveys are rarely feasible in practice. The substantial cost savings make this trade-off worthwhile, offering timely, actionable insights that support effective decision-making in resource-constrained settings. If successfully implemented, the approach could reduce reliance on extensive, resource-intensive ground surveys, enabling granular monitoring of key food security indicators with fewer financial and logistical constraints. This is particularly important in the current context of shrinking budgets for field operations and data systems, where more efficient yet reliable monitoring strategies are urgently needed. The proposed methodology is broadly applicable to other settings where timely detection of deteriorating conditions is critical and monitoring relies on surveys differing in modality, frequency, and spatial resolution.

Despite its strengths, the methodology has limitations. While it captures general trends, it may not detect sudden, localised changes driven by factors not included in the model, such as climate shocks. Incorporating climate-related covariates or allowing for time-varying or location-specific coefficients could improve responsiveness to such dynamics. The current temporal resolution (monthly) could be enhanced to weekly to enable faster detection of changes, although this would increase computational demands. In addition, scaling the analysis to include multiple years of F2F survey data would require more efficient computational strategies. Prior work, such as \cite{gellar2023calibrated}, suggests that incorporating even small amounts of F2F survey data can substantially improve prediction accuracy. Assumptions of static poststratification weights over time also warrant review, as shifts in population distributions could affect estimates.

To assess generalisability and guide broader adoption in real-world settings, there are several promising directions for future work based on the data availability of potential target countries. First, a natural next step is to conduct a pilot study in a country with a data ecosystem similar to Zimbabwe, allowing further validation and refinement of the methodology in a comparable context, and assessing its robustness and scalability. Second, we propose testing the methodology in settings where both mobile phone and F2F surveys are designed for higher administrative level analysis. In such cases, the focus would shift towards optimising the joint modelling framework for real-time small area estimation without relying on direct survey data at the desired spatial granularity, instead leveraging auxiliary sources for poststratification. Third, adapting the approach for middle-income countries, where mobile phone ownership is less correlated with socio-economic status and sampling bias from remote surveys is minimal, would allow isolation and assessment of the impact of modality bias, and demonstrate the flexibility of the jMRP approach in populations with higher digital inclusion. Finally, a promising avenue is to explore the transferability of calibration terms, specifically correction terms for modality bias, derived from well-studied countries to settings where only high-frequency mobile phone data are available. In such cases, rigorous validation and plausibility checks, incorporating expert domain knowledge, would be essential to ensure the reliability of model outputs. Exploring these pathways will help establish the broader applicability and operational utility of the proposed methodology across diverse socio-economic environments and data availability. 

\section{Acknowledgement}
This research was supported by the EPSRC (EP/V002910/2).
Elizaveta Semenova acknowledges support in part by the AI2050 program at Schmidt Sciences (G-22-64476). 

\section{Ethics approval}
This research received ethical approval from the University of Oxford, Computer Science Departmental Research Ethics Committee (DREC), on behalf of the Social Sciences and Humanities Inter-divisional Research Ethics Committee (IDREC) (CS\_C1A\_23\_011).

\section{Data availability}
The household survey data described in this study cannot be made publicly available in an open repository due to the sensitive nature of the data. Requests for household-level mVAM data can be directed to the World Food Programme (WFP), and requests for ZimVAC data can be made to the Food and Nutrition Council of Zimbabwe. Aggregated mVAM data are available via the WFP's HungerMap platform at \url{https://hungermap.wfp.org/} and can be accessed through the platform's API.
We utilised the 2015 Zimbabwe Demographic and Health Survey (DHS) and the 2019 Multiple Indicator Cluster Survey (MICS), both of which were accessed under license for this study. These data are available from their respective providers upon request: DHS data can be obtained via \url{https://dhsprogram.com/data}, and MICS data via \url{https://mics.unicef.org/}, subject to approval.
The Stan and R code used for model implementation and analysis is available at \url{https://github.com/MLGlobalHealth/Realtime-SAE-Zimbabwe}.

\newpage

\begin{appendices}
\section{Material}
\subsection{mVAM survey and  RTM system}\label{appendix: mVAM data processing}
The Mobile Vulnerability Analysis and Mapping (mVAM) program, established by the World Food Programme (WFP) in 2013, leverages mobile-phone-based surveys to monitor food security situations. Following successful pilot studies and validation of Computer-Assisted Telephone Interviewing (CATI) for food security indicators such as the Food Consumption Score (FCS) and the reduced Coping Strategy Index (rCSI), mVAM evolved into a continuous Real-Time Monitoring (RTM) system. This system collects daily data via mobile surveys, which are processed and visualized through platforms such as HungerMap \citep{HungerMap_WFP}. Additionally, the data inform predictive models for food security estimation in non-surveyed areas \citep{martini2022machine} and are used for forecasting \citep{herteux2024forecasting}. The mVAM is often used for real-time monitoring of national and sub-national trends but can also be implemented at the second or third administrative level based on country-specific needs.

Each mVAM survey follows a structured design to ensure reliability. In line with the Integrated Food Security Phase Classification (IPC) technical manual \citep{IPCmanual_31}, WFP aims to collect data from as close to 150 households per administrative area per analysis window—typically 30 or 60 days—as possible to achieve a high level of reliability. In Zimbabwe, a 30-day rolling analysis window is used, meaning that reported estimates reflect data collected within the previous 30 days. The initial sampling method is Random Digit Dialing (RDD), ensuring broad demographic and geographic coverage. Over time, WFP transitions to a panel survey design, where 80\% of respondents are pre-selected and followed for 8–10 months, while 20\% are recruited through ongoing RDD. This hybrid approach improves representativeness while maintaining sample diversity.

poststratification is applied to mitigate potential biases in the mobile-phone survey data. Households are assigned survey weights, incorporating both socio-demographic and geographic adjustments. Socio-demographic weights are derived by comparing key variables, such as education level and water source, against recent nationally representative surveys, such as the Demographic and Health Survey (DHS). These adjustments ensure that underrepresented groups are properly accounted for at the strata level. Geographic weights, updated daily, adjust for population distribution shifts and sample quotas within each administrative area. Finally, a new combined household weight is generated each day by integrating both socio-demographic and population weights, reducing the impact of sampling bias and improving the accuracy of estimates. For further details on mVAM and the RTM system, see \cite{worldfoodprogramme2021real}.

\subsection{Food Consumption Scores}\label{appendix: FCS}
The Food Consumption Score (FCS) is one of the World Food Programme's key food security indicators, designed to assess household dietary diversity and food access. It is calculated as a nutrient-weighted sum of the weekly frequency (0–7) with which a household consumes foods from eight major food groups over the past seven days, resulting in a score ranging from 0 to 112. These food groups include staples (weight: 2), pulses (3), vegetables (1), fruits (1), meat, fish, and eggs (4), dairy (4), fats (0.5), and sugar (0.5). For details on the specific food items within each group and the rationale behind the assigned weights, see \cite{worldfoodprogrammevulnerabilityanalysisandmappingbranchFoodConsumptionAnalysis2008,wfp-fcs-guidance-2024}.

FCS thresholds are used to classify households into food consumption groups, guiding the allocation of food assistance to areas and populations experiencing high levels of poor or borderline food consumption. While the standard thresholds classify households with an FCS of 21 or below as showing poor food consumption patterns and those scoring between 21 and 35 as showing borderline food consumption patterns, context-specific adjustments are sometimes applied based on dietary patterns. In Zimbabwe, where oil and sugar consumption is relatively high, the raised thresholds are used: households with an FCS of 28 or below are classified as having poor food consumption, while those scoring between 28 and 42 are categorized as having borderline food consumption.

\subsection{ZimVAC}\label{appendix: ZimVAC}
\begin{figure}
    \centering
    \includegraphics[width=0.9\linewidth]{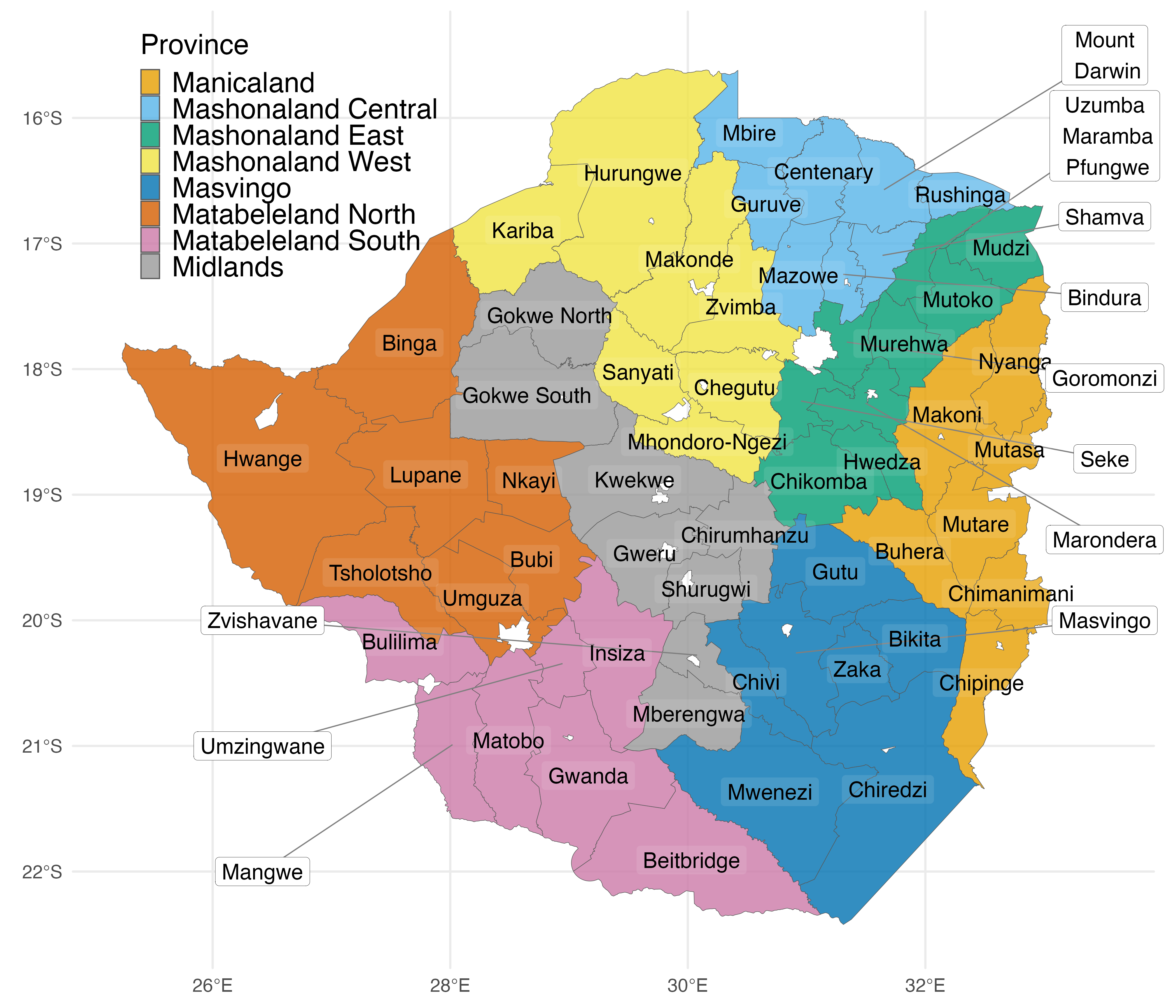}
    \caption{Map of Zimbabwe displaying the geographical boundaries of districts and provinces in Zimbabwe included in the ZimVAC 2023 and 2024 surveys}
    \label{figure: zimbabewe map}
\end{figure}

The Zimbabwe Vulnerability Assessment Committee (ZimVAC), chaired by the Food and Nutrition Council (FNC), is a multi-stakeholder consortium established in 2002 to support evidence-based responses to food security and malnutrition in Zimbabwe. The annual Rural ZimVAC survey provides key insights into livelihoods and food security conditions in rural areas to inform policy and intervention planning. The survey is designed to be statistically representative at the district, province, and national levels, with household food security as a primary indicator. The 2023 survey followed a two-stage cluster sampling approach, selecting 25 enumeration areas (EAs) per district using probability proportional to size sampling, and selecting 10 households per EA with systematic random sampling, totalling 1,500 EAs and approximately 15,000 households. From 2024, ZimVAC was renamed the Zimbabwe Livelihoods Assessment Committee (ZimLAC), with an expanded sampling scheme of 30 EAs per district, increasing the sample size to 18,000 households. The map in Figure \ref{figure: zimbabewe map} displays the eight provinces and 60 rural districts of Zimbabwe included in the Rural ZimVAC 2023 and 2024 surveys.

\section{Additional results}\label{appendix: additional results}

\subsection{Time series plots for all regions}\label{appendix:FCS time series}
\begin{figure}[ht!]
    \centering
    \includegraphics[width=0.95\textwidth]{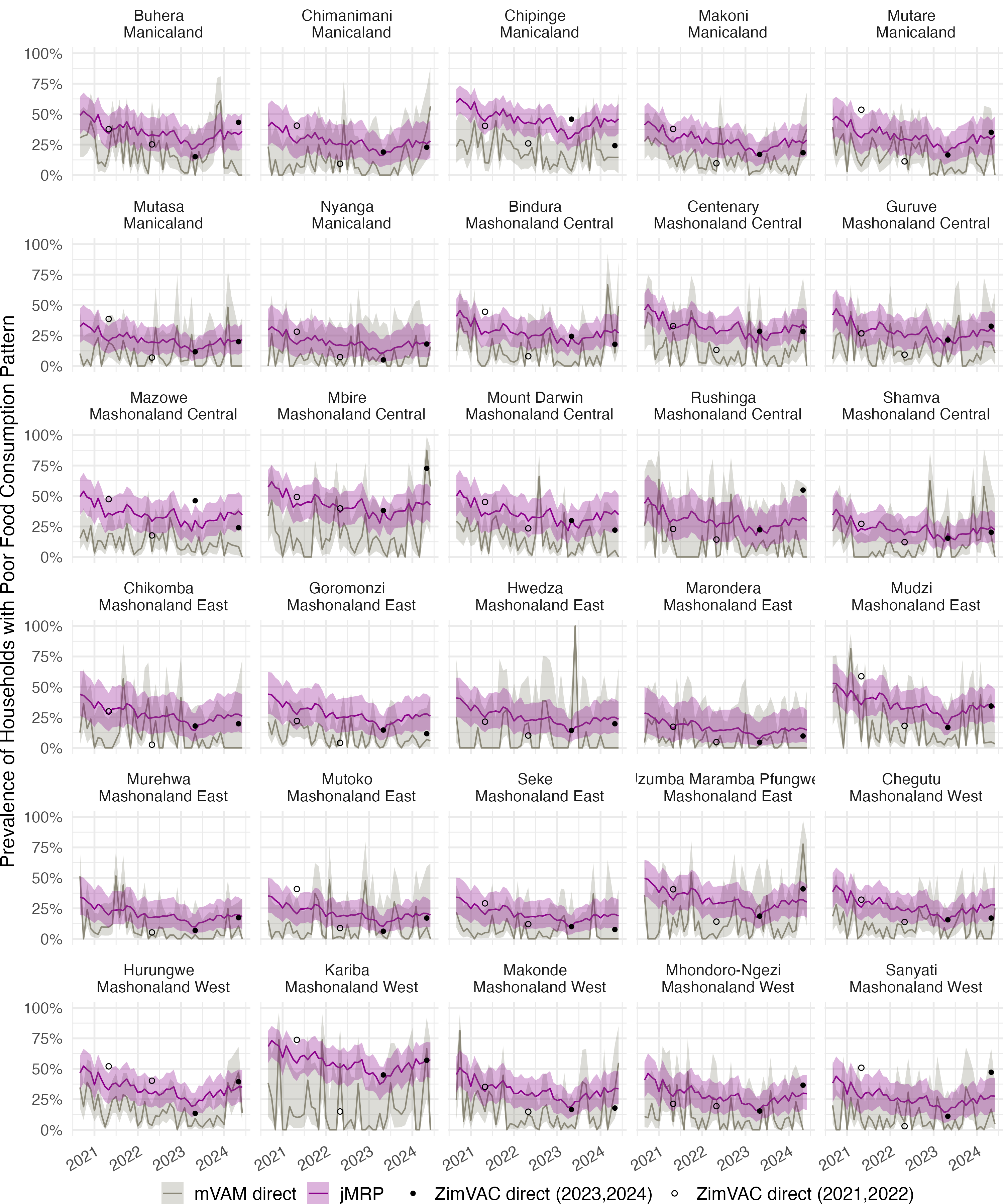}
    \caption{Prevalence of households exhibiting poor food consumption patterns (Sept 2020 - June 2024) for rural districts in the 2023 and 2024 ZimVAC survey. This figure compares food consumption estimates and their uncertainty intervals: 90\% confidence intervals for direct mobile-phone-based (mVAM) estimates (gray), 90\% credible intervals for joint MRP (ZimVAC) estimates (magenta), and point estimates for ground truth ZimVAC data (black dots). Reflecting changes in the food items for constructing the FCS index, ZimVAC estimates for 2021 and 2022 are provided for reference only. This plot shows half of the districts, with the remainder displayed in Figure \ref{figure:time trend all 2}.}
    \label{figure:time trend all 1}
\end{figure}

\begin{figure}[ht!]
    \centering
    \includegraphics[width=0.95\textwidth]{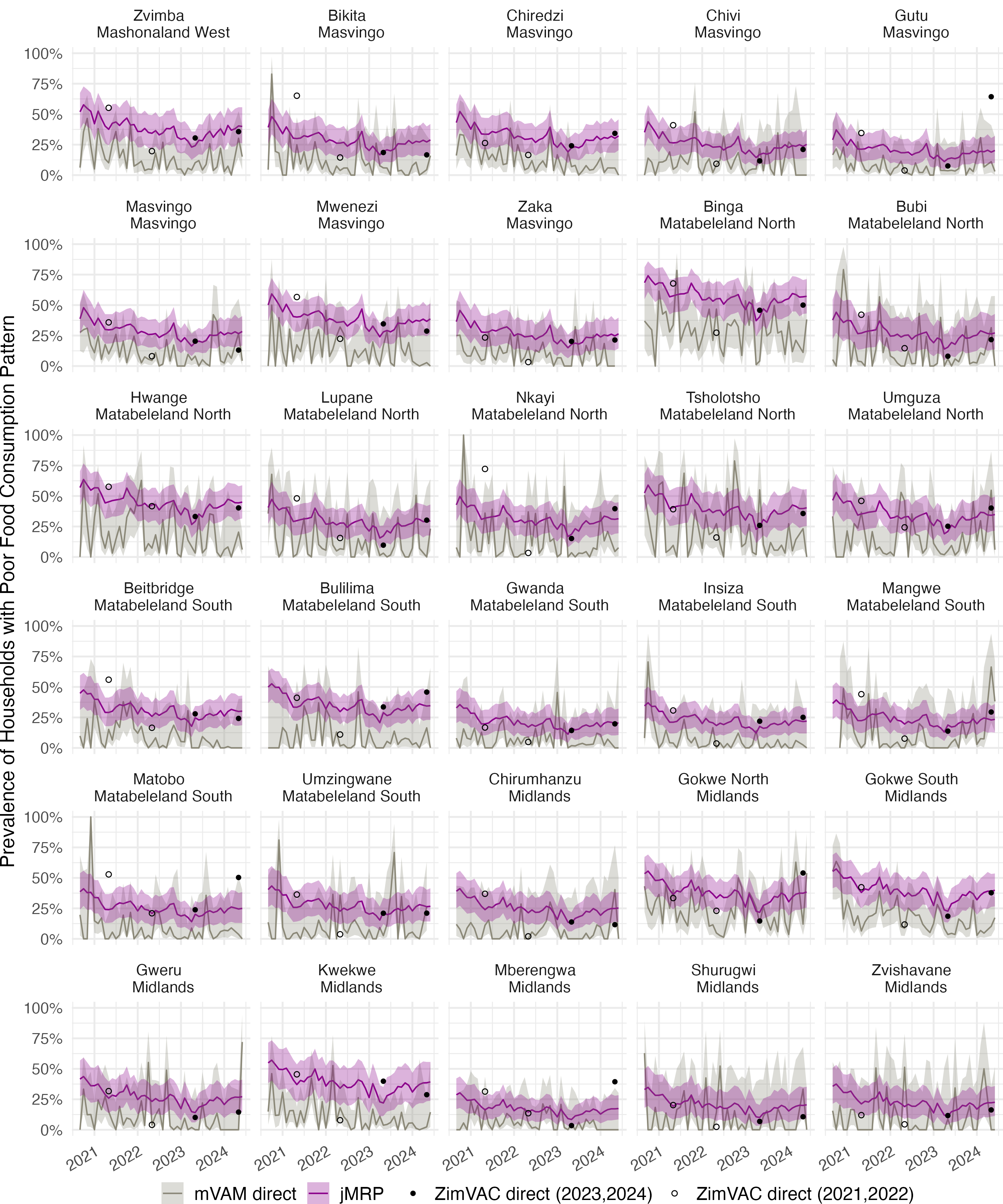}
    \caption{Prevalence of households exhibiting poor food consumption patterns (Sept 2020 - June 2024) for rural districts in the 2023 and 2024 ZimVAC survey. This figure compares food consumption estimates and their uncertainty intervals: 90\% confidence intervals for mobile-phone-based (mVAM) direct estimates (gray), 90\% credible intervals for joint MRP (ZimVAC) estimates (magenta), and point estimates for ground truth ZimVAC data (black dots). Reflecting changes in the food items for constructing the FCS index, ZimVAC estimates for 2021 and 2022 are provided for reference only. This plot shows half of the districts, with the remainder displayed in Figure \ref{figure:time trend all 1}.}
    \label{figure:time trend all 2}
\end{figure}

We present the estimated prevalence of households with poor food consumption patterns (FCS $\leq$ 28) over the period from September 2020 to June 2024 for all rural districts surveyed in the 2023 and 2024 ZimVAC. The time series plots are provided in Figures \ref{figure:time trend all 1} and \ref{figure:time trend all 2}. These plots display the trends for the entire set of districts, with mVAM direct estimates and joint MRP (jMRP) estimates compared to the ground truth ZimVAC direct estimates.

The jMRP estimates tend to be smoother and exhibit higher prevalence rates than the mVAM direct estimates, which is expected due to the joint modeling approach with poststratification, which accounts for both spatial and temporal dependencies and removes the sources of sampling and modality bias. Moreover, the jMRP estimates generally align more closely with the ZimVAC direct estimates, which are considered the ground truth for poor food consumption prevalence in the districts. This alignment further validates the effectiveness of the jMRP model in providing reliable estimates for poor food consumption prevalence at the district level over the study period. 

\subsection{Sensitivity analysis}\label{appendix: sensitivity}
The sensitivity analysis compares three different model specifications for the joint MRP (ZimVAC) estimates. The results shown in Table \ref{table: sensitivity} indicate that the choice of prior distributions (PC priors in Setting 1 vs. half-Cauchy priors in Setting 2) does not lead to significant differences in model performance, as seen in the near-identical correlation, error, and uncertainty metrics across both settings. However, allowing for district-level variation in temporal trends instead of province-level variation (Setting 3) shows notable improvements in in-sample performance (May 2023), with all evaluation metrics, including correlation coefficients, error metrics, and uncertainty quantification, improving over Setting 1. For out-of-sample performance (May 2024), Setting 3 leads to slight improvements in Pearson and Spearman correlations but does not substantially improve CCC, MAE, or CRPS. Additionally, it exhibits a higher mean bias error (MBE). It produces more conservative uncertainty quantification, as indicated by wider confidence intervals and the coverage probability showing a larger deviation (higher) from the nominal coverage. 
\begin{table}[t]
\caption{Comparison of district-level poor food consumption prevalence estimates from the joint MRP (ZimVAC) model under three different settings for May 2023 and May 2024. Performance metrics include Pearson and Spearman correlation coefficients, Lin’s Concordance Correlation Coefficient (CCC), root Mean Squared Error (rMSE), Mean Absolute Error (MAE), Mean Bias Error (MBE), 90\% coverage probability, 90\% confidence interval (CI) length, and Continuous Ranked Probability Score (CRPS). Setting 1 corresponds to the standard joint model described in Section \ref{section: model}, where the structured spatial term ($\phi_s$) follows an Intrinsic Conditional Auto-Regressive (ICAR) prior, and the structured temporal term ($\nu_t$) follows a random walk (RW) prior. The unstructured spatial ($\zeta_s$) and temporal ($\xi_t$) effects, along with the spatio-temporal interaction term ($\psi_{rt}$), which models province-level time trend variations, are assigned independent normal priors. All variance parameters are assigned Penalized Complexity (PC) priors with the specification $Pr(\sigma_u > 1) = 0.01$. Setting 2 follows the same model structure as Setting 1 but replaces the PC priors with half-Cauchy priors, using a scale parameter of 1. Setting 3 is also similar to Setting 1 but differs in that the spatio-temporal interaction term ($\psi_{st}$) models time trend variations at the district level instead of the province level.}
\label{table: sensitivity}
\centering
\begin{tabular}{lcrrrrrrrrr}
\hline
Month & Setting & \multicolumn{1}{l}{Pearson} & \multicolumn{1}{l}{Spearman} & \multicolumn{1}{l}{CCC} & \multicolumn{1}{l}{rMSE} & \multicolumn{1}{l}{MAE} & \multicolumn{1}{l}{MBE} & \multicolumn{1}{l}{Coverage} & \multicolumn{1}{l}{CI length} & \multicolumn{1}{l}{CRPS} \\ \hline
May 2023 & 1 & 0.901 & 0.888 & 0.836 & 0.053 & 0.039 & \textless{}0.001 & 0.983 & 0.195 & 0.028 \\
 & 2 & 0.901 & 0.887 & 0.836 & 0.053 & 0.049 & \textless{}0.001 & 0.983 & 0.195 & 0.028 \\
 & 3 & 0.995 & 0.990 & 0.984 & 0.018 & 0.014 & \textless{}0.001 & 1.000 & 0.169 & 0.016 \\ \hline
May 2024 & 1 & 0.534 & 0.549 & 0.474 & 0.122 & 0.091 & 0.005 & 0.917 & 0.262 & 0.066 \\
 & 2 & 0.537 & 0.557 & 0.475 & 0.121 & 0.091 & 0.006 & 0.917 & 0.263 & 0.066 \\
 & 3 & 0.579 & 0.575 & 0.476 & 0.117 & 0.090 & 0.010 & 0.933 & 0.319 & 0.065 \\ \hline
\end{tabular}
\end{table}

\end{appendices}

\newpage
\bibliographystyle{apalike}
\bibliography{reference}
\end{document}